\documentclass[aps,prl,twocolumn,superscriptaddress,10pt]{revtex4-2}
\usepackage[utf8]{inputenc}
\usepackage{amsmath, amssymb, amsfonts}
\usepackage{graphicx}
\usepackage[colorlinks=true, linkcolor=blue, citecolor=blue, urlcolor=blue]{hyperref}
\usepackage{subcaption}
\usepackage{ragged2e} 
\usepackage{tabularx} 
\usepackage{multirow} 
\usepackage{makecell}

\begin{document}

\title{Efficient direct quantum state tomography using fan-out couplings}

\author{Jaekwon Chang}
\affiliation{Department of Physics, Korea University, Seoul 02841, South Korea}

\author{Guedong Park}
\affiliation{NextQuantum Innovation Research Center, Department of Physics and Astronomy, Seoul National University, Seoul 08826, South Korea}

\author{Hyunseok Jeong}
\affiliation{NextQuantum Innovation Research Center, Department of Physics and Astronomy, Seoul National University, Seoul 08826, South Korea}

\author{Yong Siah Teo}
\email{ys\_teo@sejong.ac.kr}
\affiliation{NextQuantum Innovation Research Center, Department of Physics and Astronomy, Seoul National University, Seoul 08826, South Korea}
\affiliation{Department of Quantum Information Science and Engineering, Sejong University, Seoul 05006, South Korea}

\author{Yosep Kim}
\email{yosep9201@gmail.com}
\affiliation{Department of Physics, Korea University, Seoul 02841, South Korea}

\date{\today}

\begin{abstract}
Characterizing quantum states is essential for validating quantum devices, yet conventional quantum state tomography becomes prohibitively expensive as system size grows. Direct tomography offers a distinct route by enabling selective access to individual complex density-matrix elements, with a particular advantage for sparse target states and some verification tasks. Here we introduce a direct quantum state tomography scheme combining strong-measurement estimation with a fan-out coupling architecture. It enables mutually commuting interactions between system qubits and a single meter qubit, thereby achieving constant circuit depth, independent of system size.
Notably, the involutory fan-out coupling reduces to the identity under repetition, enabling straightforward noise scaling for quantum error mitigation. We experimentally validate the scheme on a superconducting quantum processor via the IBM Quantum Platform, demonstrating four-qubit state reconstruction and single-circuit GHZ-state fidelity estimation up to 20 qubits with error mitigation. Consistent results with standard tomography and improved efficiency establish our scheme as a promising approach to reconstructing full quantum states and scalable verification tasks.
\end{abstract}

\maketitle
\noindent{\fontsize{11.2}{13}\selectfont\textbf{Introduction}}

\noindent Quantum state tomography lies at the core of quantum characterization and verification by providing complete state information~\cite{hashim2025practical}. However, faithful reconstruction requires informationally complete data, resulting in exponential measurement overhead and substantial classical post-processing costs with increasing system size~\cite{aditi2025rigorous}. To mitigate these overheads, \textit{a priori} information about the target, such as sparsity~\cite{patel2026selective,li2024efficient} or low rank~\cite{gross2010quantum,flammia2012quantum,cramer2010efficient}, is often leveraged. In addition, adaptive measurement~\cite{huszar2012adaptive,mahler2013adaptive,kim2020universal} and machine-learning-assisted approaches have emerged as viable directions~\cite{torlai2018neural,quek2021adaptive,teo2021benchmarking,cha2021attention}. For more limited objectives, sampling-efficient verification protocols have been developed for some state classes~\cite{anshu2024survey}, including direct fidelity estimation~\cite{flammia2011direct,park2026sample} and shadow tomography~\cite{huang2020predicting,huang2025certifying,park2025resource}.

Direct quantum state tomography (DQST) provides a unified framework for both full state reconstruction and verification tasks. It enables the estimation of individual complex density-matrix elements without requiring full state reconstruction. This element-selective strategy is particularly advantageous for sparse quantum states and well suited to verification tasks, including fidelity estimation~\cite{kim2018direct}, entanglement witness~\cite{guhne2009entanglement,guhne2010separability}, and coherence measure~\cite{streltsov2017colloquium}. Moreover, on platforms where switching measurement settings is more costly than increasing measurement shots~\cite{lubinski2024optimization}, DQST can outperform randomized-measurement-based verification protocols by targeting specific matrix elements with minimal settings.

Early direct tomography was introduced through the weak-value framework based on sequential measurements~\cite{lundeen2011direct,kocsis2011observing,kim2021observing,malik2014direct,zhou2021direct}. When the first measurement is sufficiently weak, the disturbance to the system remains minimal, such that otherwise incompatible sequential measurements can still provide meaningful information~\cite{dressel2014understanding,kim2018direct}. For example, a system in $|\psi\rangle$ is weakly coupled to a meter measuring the position observable $|x\rangle\langle x|$ and subsequently post-selected in a momentum state $|p\rangle$. This procedure produces a meter shift proportional to the complex amplitude of the system state $\langle p|x\rangle \langle x|\psi\rangle$~\cite{lundeen2011direct}. Although originally formulated for pure states, the framework was extended to directly extract density-matrix elements~\cite{lundeen2012procedure,thekkadath2016direct}. Nevertheless, the weak coupling transfers only limited information and suffers from large statistical noise, which later motivated strong-measurement schemes at the cost of additional settings~\cite{vallone2016strong,zhang2020direct,calderaro2018direct,pan2019direct,zou2015direct}. 

While the scalability of DQST has been demonstrated using system-specific high-dimensional interactions~\cite{malik2014direct,zhang2020direct,zhou2021direct}, extending it to general circuit-based implementations typically requires either experimentally demanding multi-controlled gates~\cite{zou2015direct,kim2022high,nguyen2024programmable} or multiple meter qubits~\cite{calderaro2018direct,pan2019direct}. In this work, we propose an experimentally scalable DQST scheme in which a single meter qubit is strongly coupled to multiple system qubits via a single fan-out gate~\cite{hoyer2005fanout}. This architecture allows the circuit depth to be compressed to a constant~\cite{lu2019global,guo2022fanout,baumer2025measurement,song2025fanout,hashim2025entanglement}, independent of system size, while providing programmable access to arbitrary subsets of the density-matrix elements. In addition, the involutory fan-out gate reduces to the identity under repetition, enabling straightforward noise scaling for quantum error mitigation~\cite{temme2017error,kandala2019error,giurgica2020digital,henao2023adaptive}. 

To benchmark its performance, we experimentally demonstrate our DQST scheme via full state reconstruction of four-qubit states on a superconducting quantum processor using the IBM Quantum Platform~\cite{ibm_quantum}. In addition, to demonstrate efficient verification, we estimate GHZ-state fidelity for up to 20 qubits using a single circuit with quantum error mitigation. Consistent results with standard tomography and improved efficiency establish our scheme as a promising approach to reconstructing full quantum states and scalable verification tasks.

\begin{figure}[t]
    \includegraphics[width=0.93\linewidth]{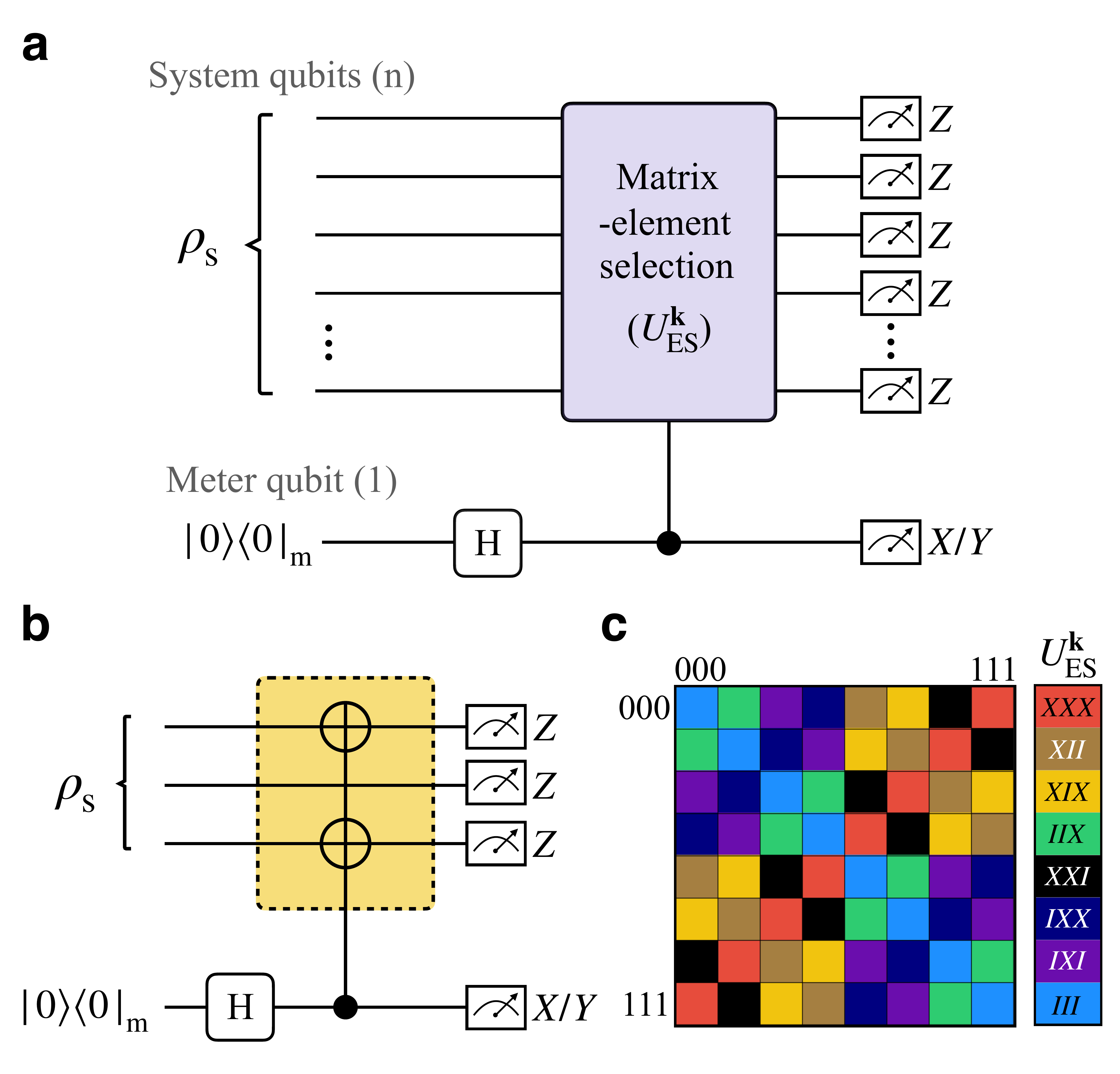}
    \caption{\justifying
    \textbf{Schematic of DQST.} \textbf{a}, Circuit diagram for matrix-element estimation of an $n$-qubit system $\rho_{\mathrm{s}}$. The meter qubit is prepared in $|+\rangle_{\mathrm{m}}$ and coupled to the system via a controlled-$U_{\mathrm{ES}}^{\mathbf{k}}$ gate. The system qubits are measured in the computational basis, while the meter qubit is measured in the $X$ and $Y$ bases to access the real and imaginary parts, respectively (see Eq.~\eqref{eq:res}).
    \textbf{b}, Example of a controlled-$U_{\mathrm{ES}}^{\mathbf{k}}$ gate with $\mathbf{k}=101$.
    \textbf{c}, Accessible matrix-element subsets for each $U_{\mathrm{ES}}^{\mathbf{k}}$. The subset corresponding to \textbf{b} is highlighted in yellow.}
    \label{fig:DQST_schematic}
\end{figure}

\vspace{1.5em}
\noindent{\fontsize{11.2}{13}\selectfont\textbf{Results}}\\
\noindent\textbf{Schematic of DQST.} Figure~\ref{fig:DQST_schematic}\textbf{a} illustrates a quantum circuit implementing our DQST scheme. A meter qubit is first prepared in $|+\rangle_\mathrm{m}=(|0\rangle_\mathrm{m}+|1\rangle_\mathrm{m})/\sqrt{2}$ using a Hadamard gate. It then interacts with an $n$-qubit target system $\rho_\mathrm{s}$ via a controlled-$U^{\mathbf{k}}_\mathrm{ES}$ gate:
\begin{align}
\lambda^{\mathbf{k}}_\mathrm{sm}
&= \frac{1}{2}\Bigl(
\rho_\mathrm{s}\otimes |0\rangle\langle 0|_\mathrm{m}
+ (U^{\mathbf{k}}_\mathrm{ES}\rho_\mathrm{s})\otimes |1\rangle\langle 0|_\mathrm{m} \label{eq:state} \\
&\quad\;
+ (\rho_{\mathrm{s}} U_\mathrm{ES}^{\mathbf{k}\,\dagger}) \otimes |0\rangle\langle 1|_\mathrm{m}
+ (U^{\mathbf{k}}_\mathrm{ES}\rho_\mathrm{s}U_\mathrm{ES}^{\mathbf{k}\,\dagger}) \otimes |1\rangle\langle 1|_{\mathrm{m}}
\Bigr).\nonumber
\end{align}
The system-meter output state $\lambda^{\mathbf{k}}_\mathrm{sm}$ shows that measuring the meter qubit in the Pauli-$X$ or $Y$ basis enables a coherent superposition of left- and right-actions of $U^{\mathbf{k}}_\mathrm{ES}$ on the system density matrix:
\begin{align}
&\langle X_{\mathbf{a}}^{\mathbf{k}}\rangle
= \mathrm{Tr}\!\left[\lambda^{\mathbf{k}}_{\mathrm{sm}}\, |\mathbf{a}\rangle\langle \mathbf{a}|_{\mathrm{s}} \otimes X_{\mathrm{m}}\right] 
= \tfrac{1}{2}\langle \mathbf{a}|
\rho_{\mathrm{s}} U_{\mathrm{ES}}^{\mathbf{k}\,\dagger}
+ U_{\mathrm{ES}}^{\mathbf{k}} \rho_{\mathrm{s}}
|\mathbf{a}\rangle
,\nonumber \\
&\langle Y_{\mathbf{a}}^{\mathbf{k}}\rangle
= \mathrm{Tr}\!\left[\lambda^{\mathbf{k}}_{\mathrm{sm}}\, |\mathbf{a}\rangle\langle \mathbf{a}|_{\mathrm{s}} \otimes Y_{\mathrm{m}}\right]
= \tfrac{i}{2}\langle \mathbf{a}|
\rho_{\mathrm{s}} U_{\mathrm{ES}}^{\mathbf{k}\,\dagger}
- U_{\mathrm{ES}}^{\mathbf{k}} \rho_{\mathrm{s}}
|\mathbf{a}\rangle,
\end{align}
where $\mathbf{a}\in\{0,1\}^n$ denotes the $n$-bit measurement outcome in the computational basis. 

To extract matrix elements of the system state, we employ a controlled-$U_\mathrm{ES}^{\mathbf{k}}$ gate such that 
$\langle \mathbf{a}|U_\mathrm{ES}^{\mathbf{k}} = \langle \mathbf{a}+\mathbf{k}|$ (mod~2), 
where $\mathbf{k}\in\{0,1\}^n$ specifies the system qubits on which Pauli-$X$ acts. 
With this choice, the real and imaginary matrix elements are obtained from meter expectation values conditioned on projection onto $|\mathbf{a}\rangle$:
\begin{align}
&\langle X_{\mathbf{a}}^{\mathbf{k}}\rangle
= \mathrm{Re}\bigl[\langle \mathbf{a}+\mathbf{k}|\rho_{\mathrm{s}}|\mathbf{a}\rangle\bigr]
= \mathrm{Re}\bigl[\langle \mathbf{a}|\rho_{\mathrm{s}}|\mathbf{a}+\mathbf{k}\rangle\bigr], \nonumber
\\
&\langle Y_{\mathbf{a}}^{\mathbf{k}}\rangle
= \mathrm{Im}\bigl[\langle \mathbf{a}+\mathbf{k}|\rho_{\mathrm{s}}|\mathbf{a}\rangle\bigr]
= -\mathrm{Im}\bigl[\langle \mathbf{a}|\rho_{\mathrm{s}}|\mathbf{a}+\mathbf{k}\rangle\bigr].\label{eq:res}
\end{align}
Sample complexity and additional details are provided in the Methods and Supplementary Note~1.

The controlled-$U_\mathrm{ES}^{\mathbf{k}}$ operation can be implemented as a fan-out gate, where a single meter qubit acts as a control and conditionally flips multiple system qubits via parallel CNOT operations. For example, the set of matrix elements $\langle \mathbf{a}|\rho_\mathrm{s}|\mathbf{a}+101\rangle$ is obtained using $U_\mathrm{ES}^{101} = X_1 I_2 X_3$, implemented via CNOT gates between the meter qubit and system qubits 1 and 3 (see Fig.~\ref{fig:DQST_schematic}\textbf{b}). Since all control-target interactions commute, they can be executed within a single circuit layer, rendering the circuit depth, in principle, independent of both the system size and the specific choice of $U^{\mathbf{k}}_\mathrm{ES}$. A single-depth fan-out gate can be experimentally implemented by simultaneously activating interactions in ion-trap and Rydberg-atom systems with all-to-all or long-range connectivity~\cite{lu2019global,guo2022fanout}. Even in superconducting qubit systems with nearest-neighbor connectivity, it can be realized at constant depth via mid-circuit measurements~\cite{baumer2025measurement,song2025fanout,hashim2025entanglement}.

\begin{figure*}[t]
    \centering
    \includegraphics[width=0.92\textwidth]{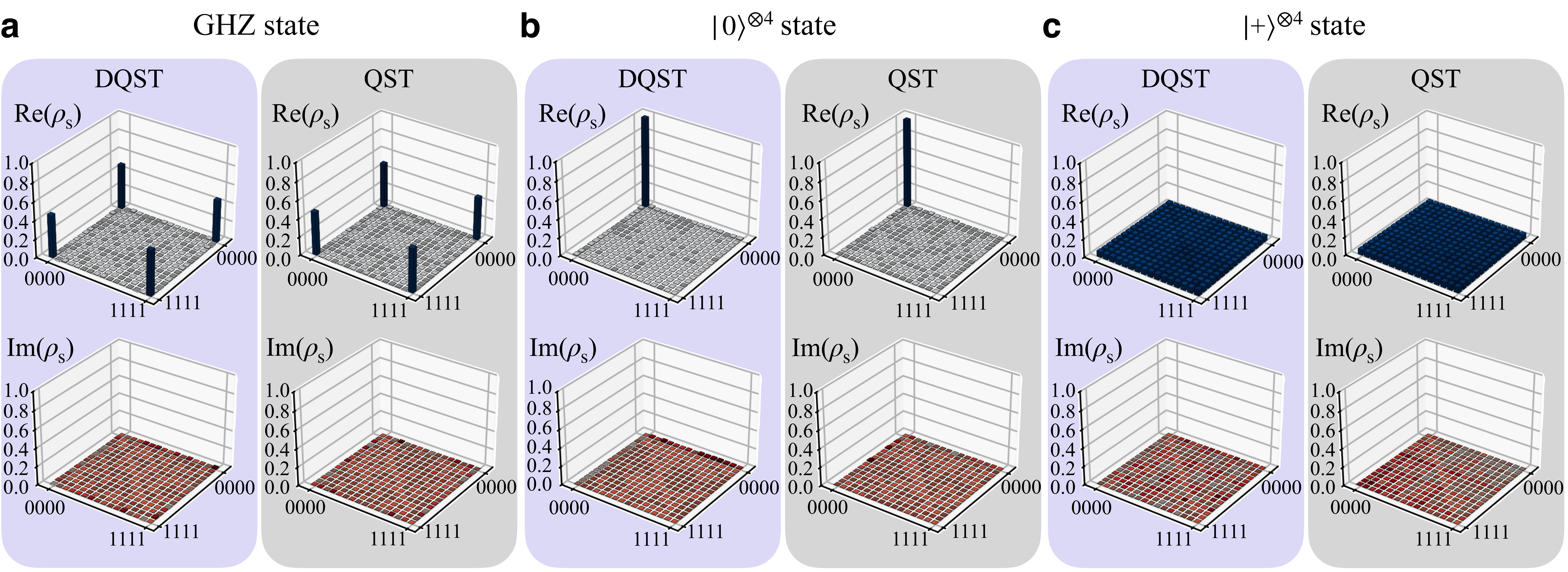}
    \caption{\justifying
    \textbf{Density matrix reconstruction results.} Four-qubit density matrices of the $|\mathrm{GHZ}_4\rangle$, $|0\rangle^{\otimes 4}$, and $|+\rangle^{\otimes 4}$ states are reconstructed via DQST (purple) and standard QST (gray). Data were collected on the \textit{ibm\_aachen} device with 10,000 shots per circuit, using 31 circuits for DQST and 81 for standard QST. Quantum readout mitigation and physical state-space projection were applied. Fidelities are reported in Table~\ref{tab:pdf_table}.
}
    \label{fig:Density matrix reconstruction result}
\end{figure*}

The choice of the matrix-element-selection operator $U^{\mathbf{k}}_\mathrm{ES}$ determines the subset of accessible elements, as illustrated in Fig.~\ref{fig:DQST_schematic}\textbf{c}. Computational basis measurement of an $n$-qubit system yields $2^n$ outcomes $\mathbf{a}$, which allow the meter measurements in Eq.~\eqref{eq:res} to access $2^n$ elements for each $\mathbf{k}$. Consequently, reconstructing the full density matrix requires $2^n$ choices of $\mathbf{k}$, each with $X$ and $Y$ meter measurements, resulting in a total of $2^{n+1}$ measurement configurations. Since the diagonal elements are real, the total number can be reduced to $2^{n+1}$-$1$. Although the scaling of DQST remains exponential in the number of system qubits $n$, this scaling is exponentially more favorable than overcomplete standard tomography, which requires $3^{n}$ of Pauli measurement settings~\cite{james2001measurement}, and compares competitively with compressed-sensing approaches that typically require $\mathcal{O}(r n^{2} 2^{n})$ settings for rank $r$~\cite{gross2010quantum,flammia2012quantum}. 

Many verification tasks require only a restricted set of density-matrix elements~\cite{kim2018direct,guhne2009entanglement,guhne2010separability,streltsov2017colloquium}. In such cases, the measurement complexity is determined by how the targeted matrix elements are distributed across the subsets accessible for each $\mathbf{k}$. If the number of relevant subsets scales polynomially with system size, the verification task can be performed efficiently. A representative example is the fidelity estimation of an $n$-qubit GHZ state, $|\mathrm{GHZ}_n\rangle = \frac{1}{\sqrt{2}}(|0\rangle^{\otimes n} + | 1\rangle^{\otimes n})$, which has four nonzero matrix elements in the computational basis. As these elements are accessible within a single DQST configuration, the fidelity can be estimated with a single measurement setting, independent of system size. Details are provided in a later subsection.

\vspace{1.5em}
\noindent\textbf{Density matrix reconstruction.} To assess the performance of DQST, we benchmark it against standard Pauli-based quantum state tomography (QST) on the \textit{ibm\_aachen} processor~\cite{ibm_quantum}.
 As the targets, we consider a 4-qubit GHZ state, $|0\rangle^{\otimes 4}$, and $|+\rangle^{\otimes 4}$, representing different levels of sparsity and entanglement. For a fair comparison, all target states are prepared on the same qubits using identical circuits for both methods, and quantum readout error mitigation (QREM) is applied~\cite{yang2022efficient}. The qubit layout is chosen to minimize the CNOT count required to implement controlled-$U_{\mathrm{ES}}^{\mathbf{k}}$ gates. While constant-depth fan-out gates are implementable on the processor~\cite{baumer2025measurement,song2025fanout,hashim2025entanglement}, we avoid their use to minimize additional experimental complexity. Details of the qubit layout, quantum circuits, and QREM are provided in Supplementary Note 2.

Figure~\ref{fig:Density matrix reconstruction result} shows density matrices reconstructed via DQST and standard QST, requiring 31 and 81 distinct circuits, respectively. To ensure physicality, they are projected onto the closest physical density matrix by minimizing the Frobenius-norm distance (see Methods). As summarized in Table~\ref{tab:pdf_table}, both methods achieve high fidelity with the ideal states, and QREM further improves the reconstruction accuracy. Notably, DQST achieves performance comparable to standard QST while using fewer than half the measurement settings. The larger statistical uncertainty in DQST arises from the smaller total shot count, as the number of shots per circuit is fixed at 10{,}000. When the total number of shots is matched, the uncertainties become comparable (see Supplementary Note 2). However, this trade-off is favorable for superconducting quantum processors, where increasing the number of shots is typically less costly than switching measurement settings~\cite{lubinski2024optimization}. 

To directly compare the two tomography methods, we evaluate the cross fidelity between the reconstructed density matrices, obtaining $98.2(2)\%$, $99.12(7)\%$, and $98.2(1)\%$ for the GHZ, $|0\rangle^{\otimes 4}$, and $|+\rangle^{\otimes 4}$ states, respectively. The discrepancies are attributed to shot noise, differences in the measurement bases, and errors in implementing controlled-$U_{\mathrm{ES}}^{\mathbf{k}}$. We also compare the raw matrix elements obtained from DQST with those of the reconstructed physical density matrix. The differences are at the level of shot noise, suggesting that post-processing can be omitted in regimes where moderate accuracy suffices. This further highlights the applicability of DQST to matrix-element-based verification tasks. Additional experimental data and analysis are provided in Supplementary Note~2.

\setlength{\tabcolsep}{4pt}
\renewcommand{\arraystretch}{1.3}
\begin{table}[t]
\centering
\begin{tabular}{c | c |c c c}
\hline\hline
 \rule{0pt}{14pt} & \makecell{Error \\ mitigation} & GHZ & $|0\rangle^{\otimes 4}$ & $|+\rangle^{\otimes 4}$ \\
\hline
\multirow{2}{*}{DQST} 
 & None & 96.3(1)\% & 98.99(6)\% & 97.57(4)\% \\
 & QREM & 97.4(1)\% & 99.30(6)\% & 98.03(5)\% \\
\hline
\multirow{2}{*}{\shortstack{Standard \\ QST}}
 & None & 96.83(4)\% & 98.75(3)\% & 97.94(4)\% \\
 & QREM & 98.02(4)\% & 99.04(3)\% & 99.20(5)\% \\
\hline\hline
\end{tabular}
\caption{\justifying
    Fidelities of the reconstructed physical density matrices in Fig.~\ref{fig:Density matrix reconstruction result} are evaluated relative to the ideal target states, both with and without quantum readout error mitigation (QREM). Uncertainties are estimated from 500 Monte Carlo resampling runs, accounting for shot noise.
} 
    \label{tab:pdf_table}
\end{table}

\begin{figure*}[ht]
    \centering
    \includegraphics[width=0.92\textwidth]{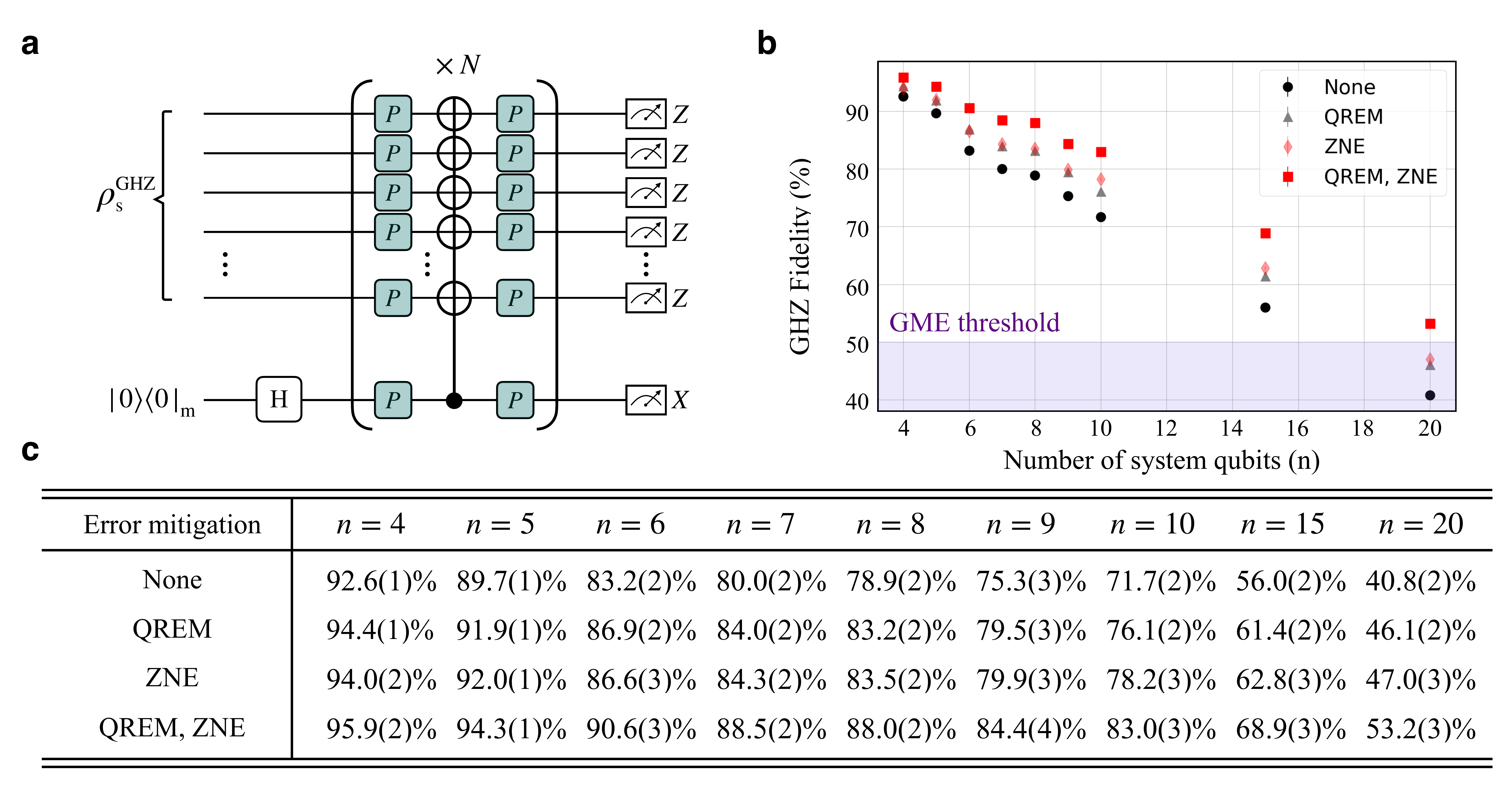}
    \caption{\small\justifying
    \textbf{GHZ-state fidelity estimation results.} \textbf{a,} Circuit diagram for direct fidelity estimation of an $n$-qubit GHZ state. To apply zero-noise extrapolation (ZNE) of the controlled-$U_{\mathrm{ES}}^{\mathbf{1}}$ gate, Pauli twirling is applied to render the noise incoherent, followed by digital gate folding with $N=1,3,5$. The zero-noise fidelity is obtained via extrapolation. \textbf{b,c,} GHZ-state fidelity as a function of qubit number under different error mitigation methods, including ZNE and quantum readout error mitigation (QREM). Each fidelity is estimated from 100,000 shots, obtained from 100 random Pauli-twirling instances with 1,000 shots each. Statistical uncertainties due to Pauli twirling are estimated via bootstrapping by resampling the 100 instances with replacement over 50 trials and are smaller than the marker size.
}
    \label{fig:direct fidelity estimation result}
\end{figure*}

\vspace{1.5em}
\noindent\textbf{GHZ-state fidelity estimation.} We now present an efficient approach to fidelity estimation of $n$-qubit GHZ states using DQST. GHZ states are central resources in quantum technologies and have long served as key benchmarks for assessing the performance of quantum platforms~\cite{javadi2025big}. However, the inefficiency of full tomography has motivated the development of alternative approaches. Direct fidelity estimation (DFE) provides a scalable framework by sampling stabilizers~\cite{flammia2011direct,park2026sample}, while parity oscillation ~\cite{guhne2007toolbox} and multiple quantum coherence (MQC) methods~\cite{wei2020verifying} extract coherence terms from interference measurements. 

Despite these advances, both DFE and parity oscillation methods require multiple distinct measurement configurations involving high-weight Pauli operators, placing stringent demands on readout fidelity and circuit reconfiguration. Although MQC alleviates some of these limitations by mapping coherence information onto the population of the $|0\rangle^{\otimes n}$ state via the inverse state-preparation unitary, it still requires 2$n$ distinct experiments and increases circuit depth. In contrast, DQST enables GHZ-state fidelity estimation using a single measurement configuration, independent of system size, requiring only projection onto $|0\rangle^{\otimes n}$. 

The GHZ fidelity is determined by four density-matrix elements corresponding to the populations and coherence between $|\mathbf{0}\rangle = |0\rangle^{\otimes n}$ and $|\mathbf{1}\rangle = |1\rangle^{\otimes n}$:
\begin{equation}
F_{\mathrm{GHZ}} = \frac{1}{2}\big(\langle \mathbf{0}|\rho_s|\mathbf{0}\rangle + \langle \mathbf{0}|\rho_s|\mathbf{1}\rangle + \langle \mathbf{1}|\rho_s|\mathbf{0}\rangle + \langle \mathbf{1}|\rho_s|\mathbf{1}\rangle\big).
\end{equation}
 The coherence terms are accessed via $U_{\mathrm{ES}}^{\mathbf{1}} = X^{\otimes n}$. Tracing out the meter after the interaction yields $\rho_\mathrm{s} + U_{\mathrm{ES}}^{\mathbf{1}}\rho_\mathrm{s} U_{\mathrm{ES}}^{\mathbf{1}\, \dagger}$ (see Eq.~\eqref{eq:state}), from which the population sum of $|\mathbf{0}\rangle$ and $|\mathbf{1}\rangle$ is directly inferred from the detection probability of $|\mathbf{0}\rangle$. Consequently, a single configuration ($\mathbf{k}=\mathbf{1}$) with an $X$-basis measurement on the meter suffices to determine the GHZ fidelity.

Figure~\ref{fig:direct fidelity estimation result}\textbf{a} shows the circuit used for fidelity estimation of an $n$-qubit GHZ state. To mitigate errors, we combine QREM with zero-noise extrapolation (ZNE) applied to the controlled-$U_{\mathrm{ES}}^{\mathbf{1}}$ operations, while leaving the target state unchanged~\cite{temme2017error,kandala2019error,giurgica2020digital}. Due to implementation constraints of $U_{\mathrm{ES}}^{\mathbf{1}}$, digital ZNE is performed at the level of individual CNOT gates~\cite{giurgica2020digital,wallman2016randomized,hashim2020randomized} (see Supplementary Note 3 for details). Figures~\ref{fig:direct fidelity estimation result}\textbf{b,c} show the measured fidelities for $n$=4\text{--}10, 15, and 20. Without mitigation, fidelities exceed the entanglement threshold of 0.5 up to 15 qubits but fall below it at 20 qubits. Applying both ZNE and QREM raises the 20-qubit fidelity above 0.5, certifying genuine multipartite entanglement (GME)~\cite{guhne2009entanglement,guhne2010separability}. Because our method uses a single measurement configuration, it reduces noise-characterization overhead compared to multi-setting approaches. Moreover, the involutory controlled-$U_{\mathrm{ES}}^{\mathbf{k}}$ is naturally compatible with pulse-level time-reversed implementations, enabling controlled noise scaling to reduce the sampling overhead of quantum error mitigation methods~\cite{henao2023adaptive,quek2024exponentially}.



\vspace{1.5em}
\noindent{\fontsize{11.2}{13}\selectfont\textbf{Discussion}}

\noindent
We have introduced and demonstrated an experimentally scalable direct quantum state tomography (DQST) scheme that combines a single meter qubit with constant-depth circuits. This approach enables full reconstruction of an $n$-qubit density matrix using $2^{n+1}$-$1$ measurement settings, while simultaneously allowing GHZ-state fidelity estimation within a single circuit configuration. By substantially reducing the number of required measurement settings, DQST provides a promising route to characterizing large-scale quantum systems and verifying genuine multipartite entanglement, with potential advantages for quantum error mitigation.

Although our experimental realization is based on a superconducting quantum processor with limited connectivity, the underlying scheme is not tied to a specific hardware platform. Architectures with high qubit connectivity, such as trapped-ion systems, naturally support direct implementations of fan-out interactions~\cite{lu2019global,guo2022fanout}, while recent advances in mid-circuit measurement and feedforward in superconducting devices provide a viable pathway to scalable realizations even in constrained connectivity settings~\cite{baumer2025measurement,song2025fanout,hashim2025entanglement}. These developments suggest that the key ingredients of DQST are already accessible across a range of quantum computing platforms.

More broadly, the structure of DQST offers a flexible framework for tailoring quantum characterization to specific tasks. By combining matrix-element selection with suitable low-cost unitary transformations, the scheme can be adapted to efficiently access relevant observables in alternative bases and to exploit sparsity in the target state. This perspective connects DQST to a wider class of measurement-efficient techniques and highlights its potential as a unifying approach to full state reconstruction and quantum verification.

\vspace{1.5em}
\noindent{\fontsize{11.2}{13}\selectfont\textbf{Methods}}

\noindent\textbf{Quantum state reconstruction.}
The density matrices obtained from DQST and standard QST may violate physicality. To obtain a valid quantum state, we project the raw estimate $\rho$ onto the set of physical density matrices by solving the convex optimization problem
$$\rho_{\mathrm{est}}
= \arg\!\min_{\sigma}\|\rho-\sigma\|_{F}
\quad
\text{s.t.}\ 
\sigma \ge 0,\;
\mathrm{Tr}(\sigma)=1,\;
\sigma=\sigma^{\dagger},
$$
where $\|\cdot\|_{F}$ denotes the Frobenius norm. This corresponds to a constrained least-squares projection in matrix space and yields the closest physical density matrix to $\rho$. Compared to maximum-likelihood estimation, which relies on an explicit noise model and iterative optimization, this approach is computationally efficient and model-independent.\\

\noindent\textbf{Sample complexity of DQST.}
Each choice of $U_\mathrm{ES}^{\mathbf{k}}$ enables the estimation of a fixed subset of density-matrix elements. For meter measurements in the $X$ and $Y$ bases (see Eq.~\eqref{eq:res}), we define the following operators associated with measurement outcome $p \in \{0,1\}$:
\begin{align}
X_\mathbf{a}^\mathbf{k}(p) 
&= \frac{1}{2}(-1)^p
\left(
|\mathbf{a}+\mathbf{k}\rangle\langle \mathbf{a}|+
|\mathbf{a}\rangle\langle \mathbf{a}+\mathbf{k}|
\right), \nonumber \\
Y_\mathbf{a}^\mathbf{k}(p) 
&= \frac{i}{2}(-1)^p
\left(
|\mathbf{a}+\mathbf{k}\rangle\langle \mathbf{a}|
-
|\mathbf{a}\rangle\langle \mathbf{a}+\mathbf{k}|
\right). \nonumber
\end{align}
These operators define unbiased estimators for the real and imaginary parts of the selected density-matrix element $\langle \mathbf{a}+\mathbf{k}|\rho|\mathbf{a}\rangle$, respectively. Since the estimators are bounded, Hoeffding’s inequality implies that estimating each matrix element to additive error $\epsilon$ with failure probability $\delta_f$ requires $\mathcal{O}\!\left(\epsilon^{-2}\log(\delta_f^{-1})\right)$ samples for a fixed measurement setting~\cite{hoeffding1963probability}. When $|K|$ distinct measurement settings are considered, applying a union bound over all settings yields a total sample complexity of $\mathcal{O}\!\left(|K|\,\epsilon^{-2}\log(|K|\,\delta_f^{-1})\right)$.\\

\noindent{\fontsize{11.2}{13}\selectfont\textbf{Data availability}}

\noindent The data that support the findings of this study are available from the corresponding author upon request.\\

\noindent{\fontsize{11.2}{13}\selectfont\textbf{Code availability}}

\noindent The code used to generate the figures within this paper and other findings of this study are available from the corresponding author upon request.

\bibliographystyle{naturemag}
\bibliography{references}

\vspace{1.5em}
\noindent{\fontsize{11.2}{13}\selectfont\textbf{Acknowledgments}}

\noindent The authors thank Jiwon Yune and Eunsung Kim for their thoughtful discussions. This work was partly supported by National Research Foundation of Korea (NRF) grant funded by the Korea government (MSIT) (RS-2024-00353348, RS-2024-00432563, RS-2025-25464760, 2020M3H3A1110365), Institute for Information \& communications Technology Planning\&Evaluation (IITP) grant funded by the Korea government (MSIT) (RS-2025-02219034), a Korea University Grant, and the faculty research fund of Sejong University in 2026. \\  

\noindent{\fontsize{11.2}{13}\selectfont\textbf{Author contributions}}

\noindent Y.K. and Y.S.T. initiated the project. Y.K., Y.S.T., and H.J. supervised the project. J.C. performed the experiments using IBM Quantum services. J.C. and G.P. carried out theoretical analysis. J.C., Y.K., and Y.S.T. wrote the manuscript with input from all authors.\\

\noindent{\fontsize{11.2}{13}\selectfont\textbf{Competing interests}}

\noindent The authors declare no competing interests.\\

\noindent{\fontsize{11.2}{13}\selectfont\textbf{Additional information}}

\noindent Correspondence and requests for materials should be addressed to  Y.S.T. and Y.K.

\newpage
\clearpage

\renewcommand{\theequation}{S\arabic{equation}}
\renewcommand{\thefigure}{S\arabic{figure}}
\renewcommand{\thetable}{S\arabic{table}}

\setcounter{page}{1}
\setcounter{figure}{0}
\setcounter{table}{0}
\setcounter{equation}{0}

\onecolumngrid
\vspace{\columnsep}
\begin{center}
\large \textbf{Supplementary Materials for \\ Efficient direct quantum state tomography using fan-out couplings}
\end{center}

\section{Supplementary Note 1 -- Direct quantum state tomography}

\subsection{A. Schematic}

We provide a more detailed description of our DQST scheme. After applying a controlled-$U_\mathrm{ES}^\mathbf{k}$ between the system and meter, the state becomes:
\begin{equation}
\lambda^{\mathbf{k}}_\mathrm{sm} = \frac{1}{2}\Bigl(
\rho_\mathrm{s}\otimes |0\rangle\langle 0|_\mathrm{m}
+ (U^{\mathbf{k}}_\mathrm{ES}\rho_\mathrm{s})\otimes |1\rangle\langle 0|_\mathrm{m}
+ (\rho_{\mathrm{s}} U_\mathrm{ES}^{\mathbf{k}\,\dagger}) \otimes |0\rangle\langle 1|_\mathrm{m}
+ (U^{\mathbf{k}}_\mathrm{ES}\rho_\mathrm{s}U_\mathrm{ES}^{\mathbf{k}\,\dagger}) \otimes |1\rangle\langle 1|_{\mathrm{m}}
\Bigr).
\label{eq:state_supp}
\end{equation}
We denote $M_\mathrm{s}$ and $M_\mathrm{m}$ as the measurement operators acting on the system and meter qubits, respectively. The measurement operators for the system qubits are defined as $M_\mathrm{s} = |\textbf{a}\rangle\langle \textbf{a}|$, where $|\textbf{a}\rangle \in \{|0\rangle, |1\rangle\}^{\otimes n}$. Consequently, for an $n$-qubit system, there exist $2^{n}$ distinct system measurement operators. For the meter qubit, four measurement operators are considered, corresponding to measurements in the $X$ and $Y$ bases: $M_\mathrm{m} \in \{ |\pm\rangle\langle \pm|,\ |\pm i\rangle\langle \pm i| \}$. Using the four measurement operators, we evaluate the corresponding detection probabilities $P_{\pm}, P_{\pm{i}}$ when the system qubits are projected onto $|\textbf{a}\rangle\langle{\textbf{a}|}$ and the meter qubit projected onto the 4 states in $M_m$. The detection probabilities are given below, where $|\textbf{a}+\textbf{k}\rangle = U_{\mathrm{ES}}^{\mathbf k} |\textbf{a}\rangle$, binary vector $\textbf{k}$.
\begin{align}
P_{\pm} &= \mathrm{Tr}(\lambda^{\mathbf{k}}_\mathrm{sm}|\textbf{a}\rangle\langle{\textbf{a}|_{\mathrm{s}}\otimes{|\pm\rangle\langle{\pm}|}_{\mathrm{m}}})=
\frac{1}{4} \big( \langle{\textbf{a}}|\rho_{\mathrm{s}}|\textbf{a}\rangle 
\pm \langle\textbf{a}+\textbf{k}|\rho_{\mathrm{s}}|\textbf{a}\rangle \pm \langle\textbf{a}|\rho_{\mathrm{s}}|\textbf{a}+\textbf{k} \rangle 
+ \langle\textbf{a}+\textbf{k}|\rho_{\mathrm{s}}|\textbf{a}+\textbf{k}\rangle \big) \nonumber \\[4pt]
P_{\pm{i}} &= \mathrm{Tr}(\lambda^{\mathbf{k}}_\mathrm{sm}|\textbf{a}\rangle\langle{\textbf{a}|_{\mathrm{s}}\otimes{|\pm{i}\rangle\langle{\pm{i}}|}_{\mathrm{m}}})=
\frac{1}{4} \big( \langle{\textbf{a}}|\rho_{\mathrm{s}}|\textbf{a}\rangle 
\mp{i} \langle\textbf{a}+\textbf{k}|\rho_{\mathrm{s}}|\textbf{a}\rangle \pm{i} \langle\textbf{a}|\rho_{\mathrm{s}}|\textbf{a}+\textbf{k} \rangle 
+ \langle\textbf{a}+\textbf{k}|\rho_{\mathrm{s}}|\textbf{a}+\textbf{k}\rangle \big)
\label{eq:p}
\end{align}

We then obtain the expectation values of X and Y:
\begin{equation}
\begin{aligned}
\langle X_{\mathbf{a}}^{\mathbf{k}}\rangle=P_{+}-P_{-}=\mathrm{Tr}(\lambda^{\mathbf{k}}_\mathrm{sm}|\textbf{a}\rangle\langle{\textbf{a}}|_{\mathrm{s}}\otimes{X_{\mathrm{m}}})=\frac{1}{2}(\langle{\textbf{a}}|\rho_{\mathrm{s}}|\textbf{a}+\textbf{k}\rangle+\langle{\textbf{a}}+\textbf{k}|\rho_{\mathrm{s}}|\textbf{a}\rangle)\\
\langle Y_{\mathbf{a}}^{\mathbf{k}}\rangle=P_{+i}-P_{-i}=\mathrm{Tr}(\lambda^{\mathbf{k}}_\mathrm{sm}|\textbf{a}\rangle\langle{\textbf{a}}|_{\mathrm{s}}\otimes{Y_{\mathrm{m}}})=\frac{i}{2}(\langle{\textbf{a}}|\rho_{\mathrm{s}}|\textbf{a}+\textbf{k}\rangle-\langle{\textbf{a}}+\textbf{k}|\rho_{\mathrm{s}}|\textbf{a}\rangle)
\end{aligned}
\label{eq:result}
\end{equation}
Since the two terms appearing in each equation are complex conjugates of one another, we may write
\begin{align}
\langle\textbf{a}+\textbf{k} |\rho_{\mathrm{s}}|\textbf{a}\rangle
&= \mathrm{Re}\bigl[\langle\textbf{a}+\textbf{k} |\rho_{\mathrm{s}}|\textbf{a}\rangle\bigr]
+ i\mathrm{Im}\bigl[\langle\textbf{a} +\textbf{k}|\rho_{\mathrm{s}}|\textbf{a}\rangle\bigr]\nonumber\\
\langle\textbf{a} |\rho_{\mathrm{s}}|\textbf{a}+\textbf{k}\rangle
&= \mathrm{Re}\bigl[\langle\textbf{a} +\textbf{k}|\rho_{\mathrm{s}}|\textbf{a}\rangle\bigr]
- i\,\mathrm{Im}\bigl[\langle\textbf{a}+\textbf{k} |\rho_{\mathrm{s}}|\textbf{a}\rangle\bigr]
\end{align}
Consequently, we arrive at Eq.~\eqref{eq:main}, which constitutes the central theoretical relation underlying our scheme.
\begin{equation}
\langle X_{\mathbf{a}}^{\mathbf{k}}\rangle= \mathrm{Re}\bigl[\langle \mathbf{a}+\mathbf{k}|\rho_{\mathrm{s}}|\mathbf{a}\rangle\bigr] ,\ \ \ \ \ \langle Y_{\mathbf{a}}^{\mathbf{k}}\rangle= \mathrm{Im}\bigl[\langle \mathbf{a}+\mathbf{k}|\rho_{\mathrm{s}}|\mathbf{a}\rangle\bigr]
\label{eq:main}
\end{equation}

Furthermore, we express the estimation of the real and imaginary parts of the density matrix using binary measurement outcomes $p \in \{+1,-1\}$ and $q \in \{+i,-i\}$, corresponding to meter measurements in the $X$ and $Y$ bases, respectively. 
\begin{align}
\sum_{\mathbf{a}} \mathrm{Re}\!\left[\langle \mathbf{a+k} | \rho_{\mathrm{s}} | \mathbf{a} \rangle\right]
\left( |\mathbf{a+k}\rangle \langle \mathbf{a}| + |\mathbf{a}\rangle \langle \mathbf{a+k}| \right)
&= \sum_{\mathbf{a}} \sum_{p}
\mathrm{Tr}\!\left[
\lambda^{\mathbf{k}}_{\mathrm{sm}}
\left( |\mathbf{a}\rangle \langle \mathbf{a}|_{\mathrm{s}} \otimes |p\rangle \langle p|_{\mathrm{m}} \right)
\right]
(-1)^{p} \nonumber \\
&\quad \times
\left( |\mathbf{a+k}\rangle \langle \mathbf{a}| + |\mathbf{a}\rangle \langle \mathbf{a+k}| \right), \\[6pt]
\sum_{\mathbf{a}} \mathrm{Im}\!\left[\langle \mathbf{a+k} | \rho_{\mathrm{s}} | \mathbf{a} \rangle \right]
\left( |\mathbf{a+k}\rangle \langle \mathbf{a}| - |\mathbf{a}\rangle \langle \mathbf{a+k}| \right)
&= \sum_{\mathbf{a}} \sum_{q}
\mathrm{Tr}\!\left[
\lambda^{\mathbf{k}}_{\mathrm{sm}}
\left( |\mathbf{a}\rangle \langle \mathbf{a}|_{\mathrm{s}} \otimes |q\rangle \langle q|_{\mathrm{m}} \right)
\right]
(-1)^{q} \nonumber \\
&\quad \times
\left( |\mathbf{a+k}\rangle \langle \mathbf{a}| - |\mathbf{a}\rangle \langle \mathbf{a+k}| \right).
\label{eq:est}
\end{align}
In other words, for a sampled outcome $(\mathbf{a},p)$, the corresponding estimator is
\[
\frac{1}{2}(-1)^{(1-p)/2}
\left(
|\mathbf{a+k}\rangle\langle \mathbf{a}| + |\mathbf{a}\rangle\langle \mathbf{a+k}|
\right),
\]
and for $(\mathbf{a},q)$,
\[
\frac{i}{2}(-1)^{(1-\mathrm{Im}[q])/2}
\left(
|\mathbf{a+k}\rangle\langle \mathbf{a}| - |\mathbf{a}\rangle\langle \mathbf{a+k}|
\right).
\]
The factor of $1/2$ arises from double counting under the exchange $\mathbf{a} \leftrightarrow \mathbf{a+k}$. Since the estimators are bounded, i.e.,
\[
\left\|(-1)^{p}\left(|\mathbf{a+k}\rangle\langle \mathbf{a}| + |\mathbf{a}\rangle\langle \mathbf{a+k}|\right)\right\|_2 \le 2,
\]
Hoeffding’s inequality~\cite{hoeffding1963probability} implies that, for a fixed $\mathbf{k}$, estimating the corresponding operator (and its imaginary counterpart) within $\epsilon$ Frobenius error requires $\mathcal{O}\!\left(\epsilon^{-2}\log(\delta_f^{-1})\right)$ samples, where $\delta_f$ denotes the failure probability.

\begin{figure*}[b]
    \centering
    \includegraphics[width=0.88\textwidth]{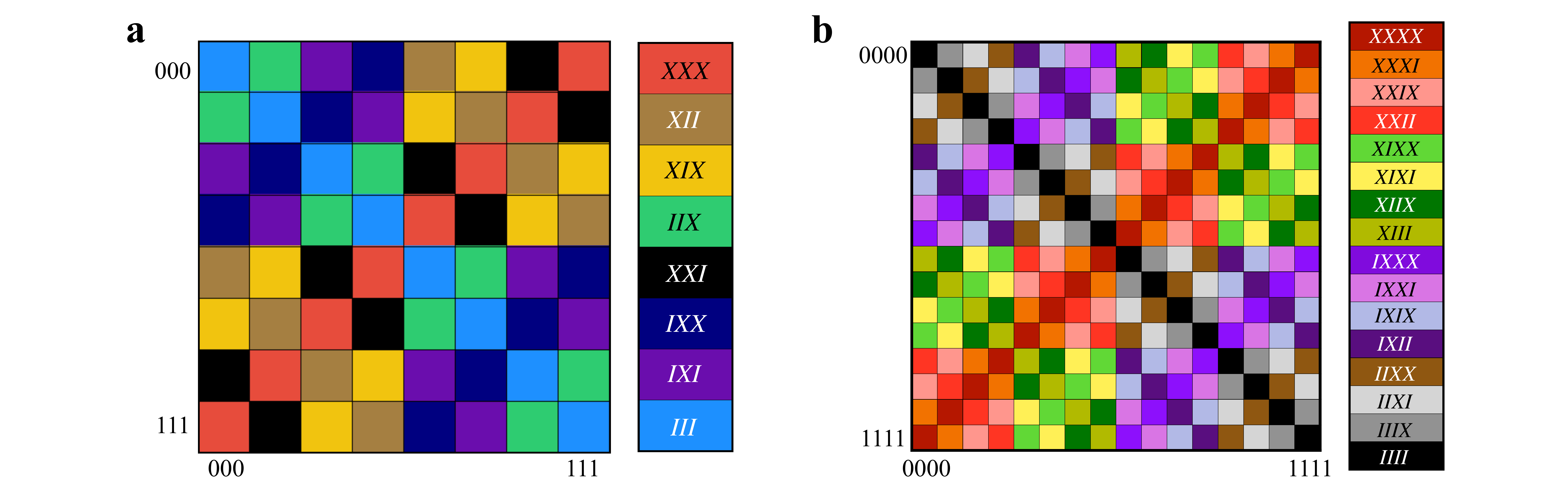}
    \caption{\small\justifying
Density matrix acquired from each configuration using DQST.  
\textbf{a,} Three-qubit case. For each configuration of $U_{\mathrm{ES}}^{\mathbf{k}} \in \{X, I\}^{\otimes 3}$, eight elements of the density matrix are directly obtained, enabling the reconstruction of the full density matrix.
\textbf{b,} Four-qubit case. The same procedure is extended to $U_{\mathrm{ES}}^{\mathbf{k}} \in \{X, I\}^{\otimes 4}$, where each configuration yields sixteen density matrix elements, allowing full state reconstruction.
}
    \label{fig:dens_34}
\end{figure*}

\begin{figure*}[bh]
    \centering
    \includegraphics[width=0.88\textwidth]{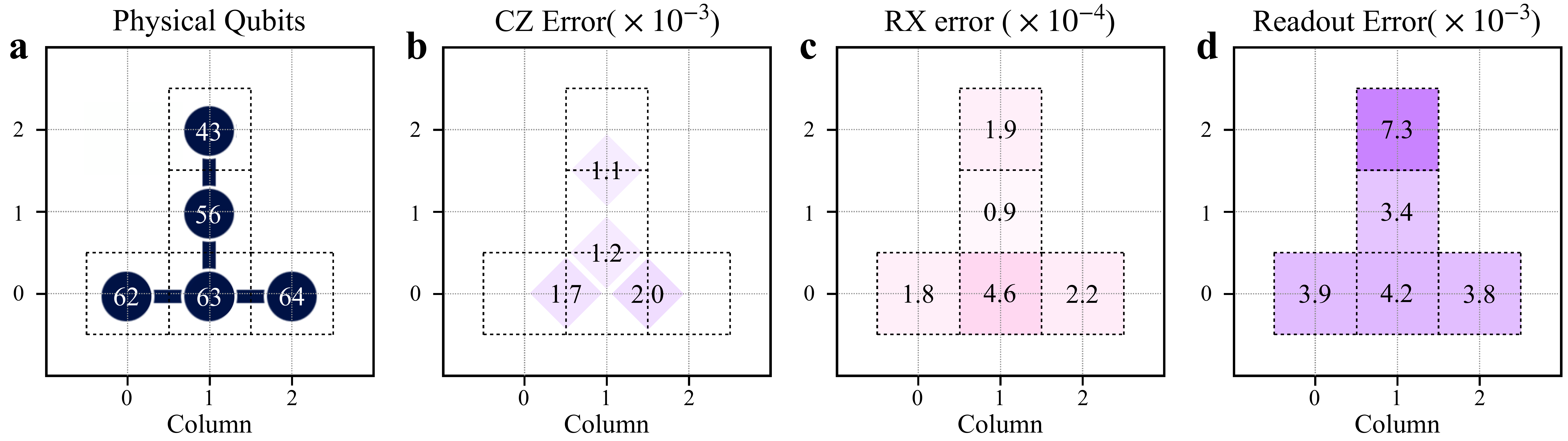}
    \caption{\small\justifying
Hardware configuration for full density-matrix reconstruction using DQST.
\textbf{a,} Physical qubits employed for the $n=4$ density-matrix reconstruction experiments. The selected qubits $\{$43, 56, 62, 64, 63$\}$ were chosen to maximize the connectivity of the designated control qubit (63). 
\textbf{b,} Two-qubit $CZ$ error rates between adjacent qubits. The average $CZ$ error rate is $1.5\times10^{-3}$. 
\textbf{c,} Single-qubit $RX$ gate error rates of the selected qubits. The average $RX$ error rate is $2.28\times10^{-4}$, showing that single qubit gate error is significantly lower than two qubit gate error. 
\textbf{d,} Readout error rates of the selected qubits. The average readout error is $4.52\times10^{-3}$, indicating that readout error is the dominant error source.
}
    \label{fig:exp_1_config}
\end{figure*}

\begin{figure*}[bh!]
    \centering
    \includegraphics[width=0.9\textwidth]{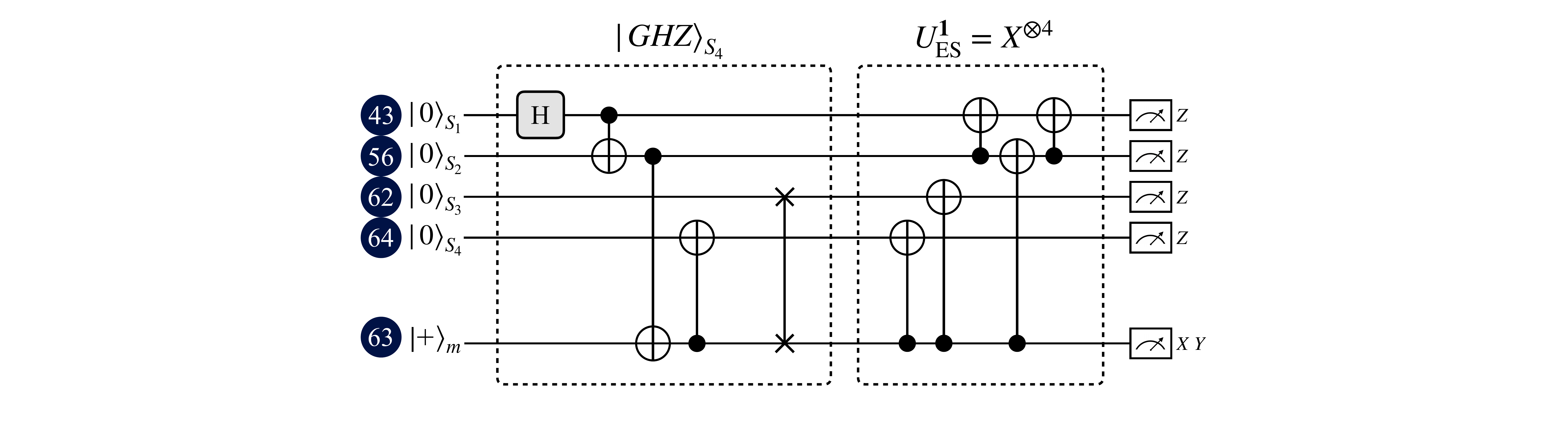}
    \caption{\small\justifying
Circuit employed for $|GHZ\rangle_{S_{4}}$ full density matrix reconstruction with $U_{\mathrm{ES}}^{\mathbf{1}}=X^{\otimes 4}$.  
For the preparation of the GHZ state, we first entangled qubits 43, 56, 63, and 64, and subsequently employed an additional SWAP gate to generate an entangled state involving qubits 43, 56, 62, and 64.  
For matrix-element selection ($U_{\mathrm{ES}}^{\mathbf{1}}=X^{\otimes 4}$), we directly applied three $CNOT$ gates with qubit 63 as the control, taking advantage of the direct connectivity. Subsequently, we applied two additional $CNOT$ gates using qubit 56 as the control and qubit 43 as the target.
     }
    \label{fig:circuit layout}
\end{figure*}

\subsection{B. Matrix-element selection}

Reconstruction of the full density matrix with our protocol requires $2^{n+1}-1$ circuit configurations, obtained by enumerating all possible matrix-element selections as described in Eq.~\eqref{eq:full_tomography}.
\begin{eqnarray}
\rho_s=\sum_{{\bf{a}}}\Bigr[&&\langle X_{\mathbf{a}}^{\mathbf{0}}\rangle|\mathbf{a}\rangle\langle\mathbf{a}|
\nonumber\\
&&+\sum_{{\bf{k\ne 0}}}\Bigr(\langle X_{\mathbf{a}}^{\mathbf{k}}\rangle(|\mathbf{a}+\mathbf{k}\rangle\langle\mathbf{a}|+|\mathbf{a}\rangle\langle\mathbf{a}+\mathbf{k}|)
+i\ \langle Y_{\mathbf{a}}^{\mathbf{k}}\rangle(|\mathbf{a}+\mathbf{k}\rangle\langle\mathbf{a}|-|\mathbf{a}\rangle\langle\mathbf{a}+\mathbf{k}|)\Bigr)\Bigr]
\label{eq:full_tomography}
\end{eqnarray}
The off-diagonal elements of the density matrix are generally complex-valued and therefore require measurements in both the $X$ and $Y$ bases of the meter qubit. This is performed for all possible configurations of $U_{\mathrm{ES}}^{\mathbf{k}}$, amounting to $2^{n}$ choices, except for the case $U_{\mathrm{ES}}^{\mathbf{0}}=I^{\otimes n}$ ($\mathbf{k}=\mathbf{0}$), where $\mathbf{0}$ denotes the all-zeros vector. As a result, a total of $2^{n+1}-2$ circuit configurations are required to access all complex off-diagonal elements.
In contrast, the diagonal elements are always real-valued and can be obtained from a single circuit configuration with $U_{\mathrm{ES}}^{\mathbf{0}}=I^{\otimes n}$ by measuring the meter qubit in the $X$ basis. Consequently, full quantum state tomography requires $2^{n+1}$-$1$ distinct circuit configurations.

For example, in the case of full tomography of a three-qubit system, there are eight possible matrix-element selection operators as shown in Fig.~\ref{fig:dens_34}\textbf{a}. The measurement outcome states of the system qubits are given by
$|\textbf{a}\rangle \in \{|0\rangle,|1\rangle\}^{\otimes 3}$. Consequently, each configuration of $U_{\mathrm{ES}}^{\mathbf{k}}$ yields eight density-matrix elements, allowing all 64 complex density-matrix elements to be reconstructed, as illustrated in Fig.~\ref{fig:dens_34}\textbf{a}. Similarly, in the four-qubit case, the use of all 16 matrix-element selection operators enables the reconstruction of all $256$ density-matrix elements, as shown in Fig.~\ref{fig:dens_34}\textbf{b}.

\section{Supplementary Note 2 -- Full state tomography}
\subsection{A. Experimental details}

We describe the experimental configurations employed on the \textit{ibm\_aachen} device for full tomography~\cite{ibm_quantum}. Figure~\ref{fig:exp_1_config} shows the physical qubits selected for the experiment together with their corresponding error characteristics. All circuit configurations were designed while explicitly accounting for the qubit connectivity constraints of the \textit{ibm\_aachen} processor. In an ideal setting, quantum fan-out operations can be implemented with constant circuit depth. However, current IBM quantum processors do not natively support single-depth fan-out gates. Consequently, fan-out operations must be decomposed into sequences of $CNOT$ gates, and additional $SWAP$ operations are required when qubits are not directly connected. Such operations inevitably increase circuit depth and noise.

To mitigate this overhead, we carefully selected adjacent, physically connected qubits and designed connectivity-aware circuits. Figure~\ref{fig:circuit layout} illustrates the circuit configuration used for DQST-based reconstruction of a four-qubit GHZ state. The physical qubits employed were $\{43, 56, 62, 64, 63\}$, corresponding to logical qubits $q_1$ through $q_5$, where $q_5$ (qubit 63) serves as the meter qubit.

Matrix-element selection requires the meter qubit to control multiple system qubits via fan-out operations implemented using $CNOT$ gates. However, each qubit on the \textit{ibm\_aachen} device is typically connected to at most three neighboring qubits. To overcome this limitation, three $CNOT$ gates were applied directly from the meter qubit $q_5$ to its nearest neighbors, while two additional $CNOT$ gates were applied with $q_2$ as the control and $q_1$ as the target, thereby propagating the fan-out operation to $q_1$ (right box in Fig.~\ref{fig:circuit layout}). This approach reduces the total number of $CNOT$ gates required to realize an effective generalized fan-out structure. All 16 matrix-element selection operators used for full density matrix reconstruction were constructed following the same qubit layout strategy.

For $|GHZ\rangle_{S_4}$ state preparation, entanglement was first generated among qubits $q_1$, $q_2$, $q_5$, and $q_4$ using directly connected qubits. Subsequently, a $SWAP$ operation was applied between $q_3$ and $q_5$ to transfer the entangled state onto qubits $q_1$ through $q_4$, thereby matching the qubit layout employed in the matrix-element selection.

\begin{figure*}[b]
    \centering
    \includegraphics[width=0.88\textwidth]{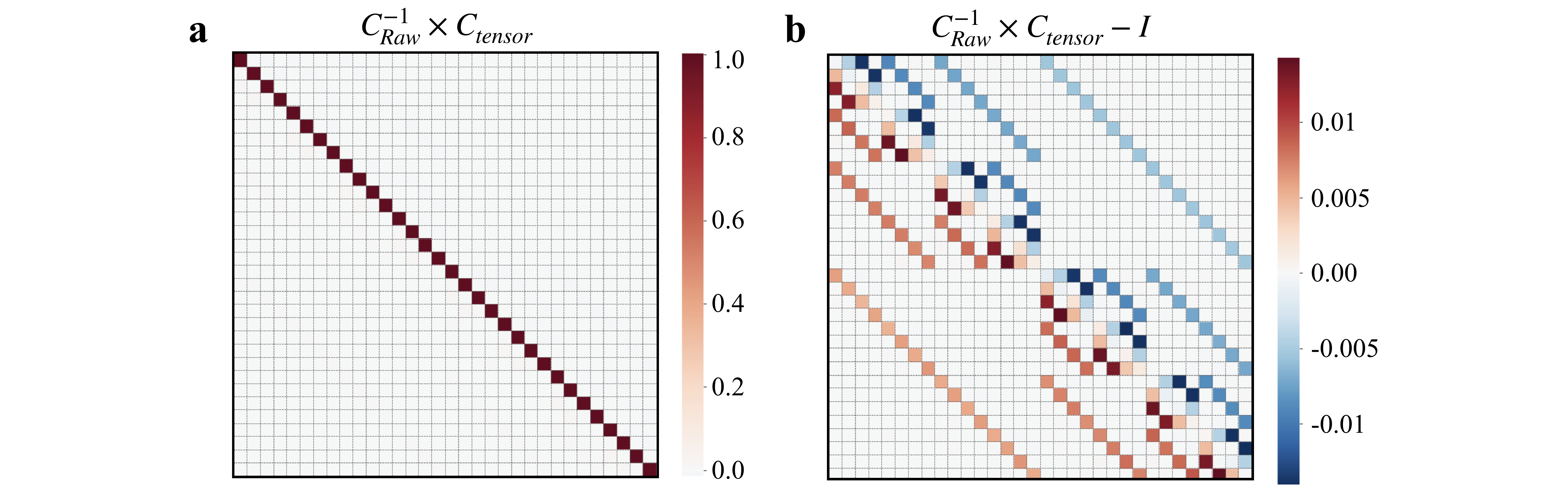}
    \caption{\small\justifying
Comparison between five qubit $C_\text{Raw}$ and $C_\text{tensor}$.
\textbf{a,} Result of $C_{\mathrm{Raw}}^{-1} C_{\mathrm{tensor}}$. If the two confusion matrices are equivalent, this product should yield the identity matrix. The result closely resembles the identity, confirming the agreement between the two matrices.  
\textbf{b,} Difference between the result in Fig.~\ref{fig:QREM}\textbf{a} and the identity matrix. 
}
    \label{fig:QREM}
\end{figure*}

\subsection{B. Readout error mitigation}

This subsection describes the readout error mitigation strategy employed in our experiments. The approach is based on constructing and inverting a confusion matrix~\cite{yang2022efficient}, which captures the conditional probabilities of readout outcomes. Specifically, each qubit is prepared in either the $|0\rangle$ or $|1\rangle$ state, and the probability of measuring $|0\rangle$ or $|1\rangle$ is recorded, yielding a $2 \times 2$ confusion matrix for a single qubit, as shown in Eq.~\eqref{eq:confmat}, where $P(b|a)$ represents the probability of measuring outcome $|b\rangle$ when the state $|a\rangle$ is prepared. For systems with a small number of qubits, the full confusion matrix could be directly characterized. However, for larger systems, direct characterization becomes impractical due to exponential scaling. In such cases, we first characterized the individual single-qubit confusion matrices and then constructed the full $n$-qubit confusion matrix as a tensor product of the individual matrices. By inverting and applying this matrix to the measured outcome distributions, the readout errors are mitigated.
\begin{equation}
C = 
\begin{bmatrix}
P(0|0) & P(0|1) \\
P(1|0) & P(1|1)
\end{bmatrix}
\label{eq:confmat}
\end{equation}

We compared the confusion matrices obtained by two different methods in order to verify that the tensor product of the individual confusion matrices ($C_{tensor}$) is equivalent to the confusion matrix derived from measuring all outcomes of the $n$-qubit system ($C_{raw}$).
\begin{equation}
C_\text{Raw} = 
\begin{bmatrix}
P(0^{\otimes n} \mid 0^{\otimes n}) & \cdots & P(0^{\otimes n} \mid 1^{\otimes n}) \\
\vdots & \ddots & \vdots \\
P(1^{\otimes n} \mid 0^{\otimes n}) & \cdots & P(1^{\otimes n} \mid 1^{\otimes n})
\end{bmatrix},\ \ C_\text{tensor} = C_1 \otimes{C_2}\otimes{C_3}...\otimes{C_n}
\label{eq:confmat_raw}
\end{equation}

We experimentally demonstrated this by calculating a five-qubit confusion matrix.
The raw confusion matrix $C_{\mathrm{Raw}}$ was obtained by preparing all computational basis states and measuring the corresponding outcome probabilities. The tensor-product confusion matrix $C_{\mathrm{tensor}}$ was constructed by first obtaining the individual $2\times 2$ confusion matrices for each qubit and then taking their tensor product.  

Figure~\ref{fig:QREM}\textbf{a} presents the result of $C_{\mathrm{Raw}}^{-1} C_{\mathrm{tensor}}$, which should ideally yield the identity matrix if the two methods are equivalent.  
Figure~\ref{fig:QREM}\textbf{b} shows the difference between this result and the identity matrix.  
Since the matrix in Fig.~\ref{fig:QREM}\textbf{a} closely resembles the identity and each entry of Fig.~\ref{fig:QREM}\textbf{b} lies within the range $[-0.02, 0.02]$, we conclude that $C_{\mathrm{tensor}}$ is in good agreement with $C_{\mathrm{Raw}}$.

\begin{figure*}[b]

    \centering
    \includegraphics[width=0.9\textwidth]{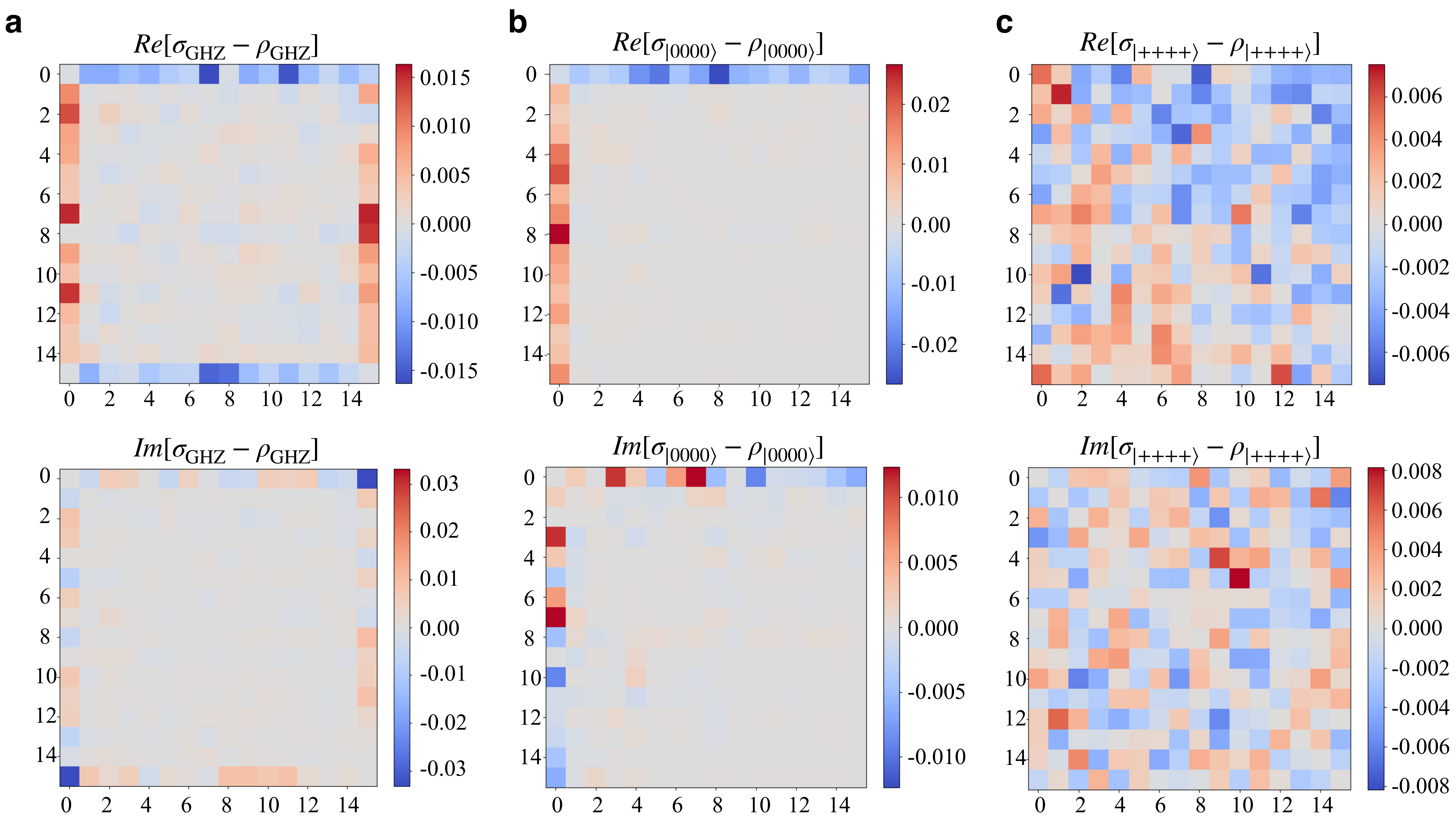}
    \caption{\small\justifying 
Element-wise differences between the raw density matrix ($\rho$) reconstructed via DQST with readout error mitigation applied and the density matrix projected onto the physical state space ($\sigma$). The top (bottom) row shows the real (imaginary) parts of the differences: \textbf{a,} $\sigma_{\mathrm{GHZ}}-\rho_{\mathrm{GHZ}}$, \textbf{b,} $\sigma_{|0000\rangle}-\rho_{|0000\rangle}$, and \textbf{c,} $ \sigma_{|++++\rangle}-\rho_{|++++\rangle}$.}
    \label{fig:state_space}
\end{figure*}

\subsection{C. State reconstruction}

The density matrix is obtained by combining $2^{n+1}-1$ distinct measurement outcomes via DQST. Subsets of the density matrix are first reconstructed independently and subsequently merged using all available measurement data. However, due to statistical noise and reconstruction inconsistencies arising from this fusion procedure, the resulting matrix does not, in general, satisfy the physical constraints required of a valid density matrix—namely, unit trace and positive semi-definiteness. To address this issue, we project the reconstructed matrix onto the space of physical states, thereby enforcing these constraints and ensuring a physically valid density operator.

Figure~\ref{fig:state_space} shows the element-wise differences between the raw density matrix ($\rho$) reconstructed via DQST (with QREM applied) and the corresponding projected state ($\sigma$) for the three states considered in our experiment. The top (bottom) row displays the real (imaginary) parts of the differences. We observe that the deviations are generally small across most matrix elements, indicating that the raw reconstruction is already close to a physical state. Larger discrepancies appear predominantly near specific elements, reflecting the correction imposed by the projection procedure to enforce positivity and unit trace.

\subsection{D. Comparison between standard QST and DQST}
We present the fidelities obtained via standard QST and DQST under the same total number of measurement shots. 
In the main text, 10,000 shots per circuit were used, resulting in different total numbers of shots for standard QST and DQST, and consequently different standard deviations. 
In contrast, Table~\ref{tab:table} reports the fidelities with the ideal states for the three target states using an equal total number of shots for both methods. 
Specifically, 3,827 shots per circuit were used for standard QST to match the total number of shots used in DQST. As shown in Table~\ref{tab:table}, matching the total number of measurement shots leads to comparable standard deviations between standard QST and DQST across all three target states, indicating that the observed differences in the main text primarily originate from unequal shot budgets.

\begin{figure*}[t]

    \centering
    \includegraphics[width=0.88\textwidth]{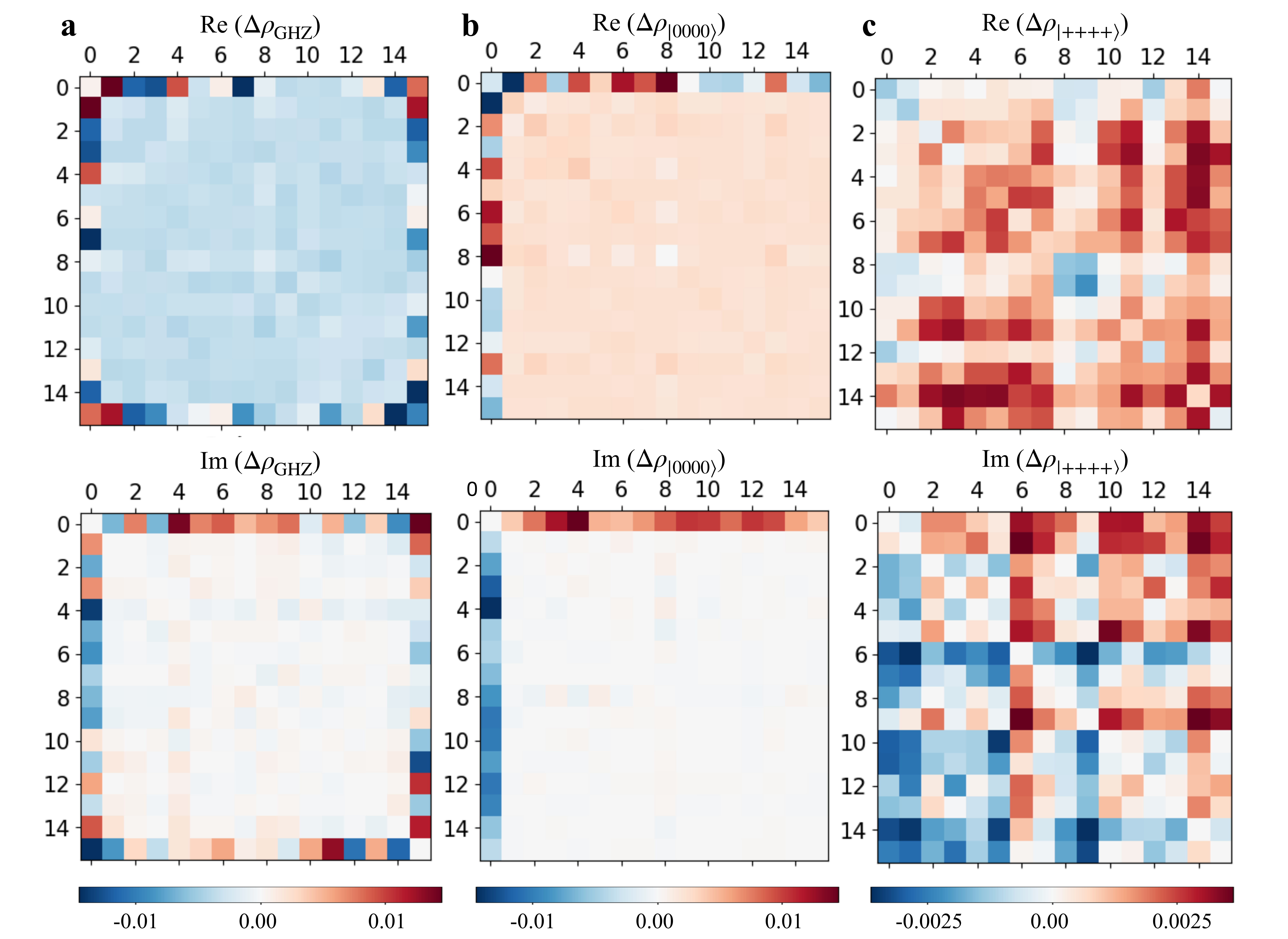}
    \caption{\small\justifying 
Difference between density matrix obtained \emph{via} standard QST and DQST.
$\Delta\rho$ of the three states we reconstructed \emph{via} standard QST and DQST. We show the values of each matrix elements of the real components of $\Delta\rho$, and the imaginary components of $\Delta\rho$ of \textbf{a,} $|{GHZ}\rangle_{S_4}$, \textbf{b,} $|0\rangle^{\otimes{4}}$, and \textbf{c,} $|+\rangle^{\otimes{4}}$.
}
    \label{fig:QSTvs.DQST}
\end{figure*}

Next, we present the differences between the reconstructed density matrices obtained via standard QST and DQST, as shown in Fig.~\ref{fig:QSTvs.DQST}.
Figure~\ref{fig:QSTvs.DQST}\textbf{a} compares the element-wise differences for the GHZ state. The observed deviations are small compared to the dominant off-diagonal elements of the ideal GHZ density matrix, indicating a high degree of agreement between the two reconstruction methods.
Figure~\ref{fig:QSTvs.DQST}\textbf{b} shows the corresponding difference for the  \(|0\rangle^{\otimes 4}\)
state, where the deviations remain negligible relative to the single non-zero diagonal element of the ideal state.
Finally, Fig.~\ref{fig:QSTvs.DQST}\textbf{c} presents the result for the \(|+\rangle^{\otimes 4}\)
state, demonstrating uniformly small discrepancies across all matrix elements. Together, these results confirm the consistency of DQST with standard QST across representative entangled and separable states.
\begin{equation}
D(\rho_\text{DQST},\rho_\text{QST})=\frac{1}{2}||\rho_\text{DQST}-\rho_\text{QST}||_1=\frac{1}{2}Tr(\sqrt{(\rho_\text{DQST}-\rho_\text{QST})^{\dagger}(\rho_\text{DQST}-\rho_\text{QST})})
\label{eq:tracedistance}
\end{equation}

To further quantify the similarity between the two reconstructed density matrices, we compute the trace distance between the results obtained from standard QST and DQST for the three target states. The trace distance is defined as in Eq.~\eqref{eq:tracedistance}. For the GHZ state, the trace distance is \(0.083\), whereas for the \(|0\rangle^{\otimes 4}\) state and the \(|+\rangle^{\otimes 4}\) state, the trace distances are \(0.063\) and \(0.051\), respectively. These results confirm that the density matrices reconstructed by DQST closely resemble those obtained by standard QST.

\begin{table}[t]
    \centering
    \includegraphics[width=0.9\textwidth]{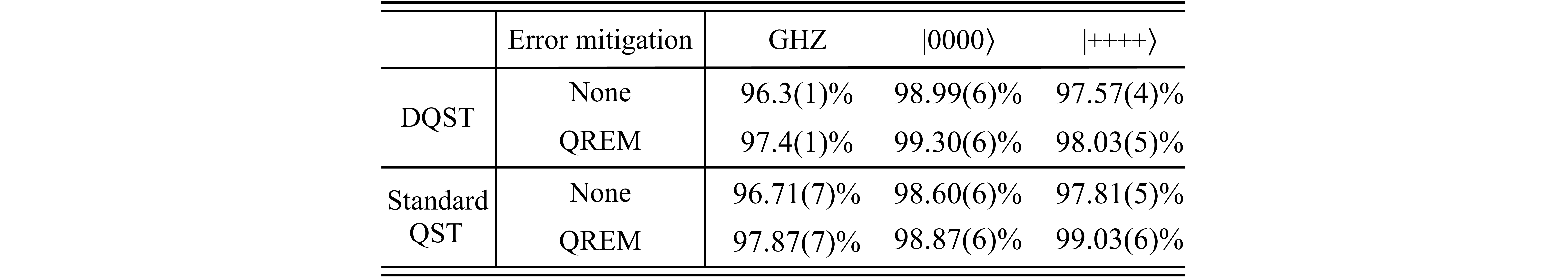}
    \caption{\small\justifying
State-reconstruction fidelities obtained \emph{via} standard QST and DQST, compared with the ideal target states. 
The total number of measurement shots was matched between the two methods by using 10,000 shots for DQST and 3,827 shots per circuit for standard QST. 
Uncertainties were estimated from 500 Monte Carlo resampling runs, accounting for shot noise.} 
    \label{tab:table}
\end{table}

\section{Supplementary Note 3 -- GHZ-state fidelity estimation}

\subsection{A. GHZ-state fidelity estimation using DQST}

As described in the main text, the fidelity of an $n$-qubit GHZ state can be estimated by employing $U_{\mathrm{ES}}^{\mathbf{1}} = X^{\otimes n}$ (with $\mathbf{k}=\mathbf{1}$ denoting the all-ones vector) and measuring the meter qubit in the $X$ basis. Choosing the computational basis states $|\textbf{a}_0\rangle = |0\rangle^{\otimes n}$ and $|\textbf{a}_1\rangle = |1\rangle^{\otimes n}$, the operation $U_{\mathrm{ES}}^{\mathbf{1}} =X^{\otimes n}$ maps them to $|\bar{\textbf{a}}_0\rangle=|\textbf{a}_0+\mathbf{1}\rangle = |1\rangle^{\otimes n}$ and $|\bar{\textbf{a}}_1\rangle = |\textbf{a}_1+\textbf{1}\rangle=|0\rangle^{\otimes n}$, respectively. Consequently, the following two equations are obtained.
\begin{align}
P_{\pm}^{0} &= \mathrm{Tr}(\lambda^{\mathbf{1}}_\mathrm{sm}|\textbf{a}_0\rangle\langle{\textbf{a}_0|_{\mathrm{s}}\otimes{|\pm\rangle\langle{\pm}|}_{\mathrm{m}}})=
\frac{1}{4} \big( \langle{\textbf{a}_0}|\rho_{\mathrm s}|\textbf{a}_0\rangle 
\pm \langle\bar{\textbf{a}}_0|\rho_{\mathrm s}|\textbf{a}_0\rangle \pm \langle\textbf{a}_0|\rho_{\mathrm s}|\bar{\textbf{a}}_0\rangle 
+ \langle\bar{\textbf{a}}_0|\rho_{\mathrm s}|\bar{\textbf{a}}_0\rangle \big) \\
P_{\pm}^{1} &= \mathrm{Tr}(\lambda^{\mathbf{1}}_\mathrm{sm}|\textbf{a}_1\rangle\langle{\textbf{a}_1|_{\mathrm{s}}\otimes{|\pm\rangle\langle{\pm}|}_{\mathrm{m}}})=
\frac{1}{4} \big( \langle{\textbf{a}_1}|\rho_{\mathrm s}|\textbf{a}_1\rangle 
\pm \langle\bar{\textbf{a}}_1|\rho_{\mathrm s}|\textbf{a}_1\rangle \pm \langle\textbf{a}_1|\rho_{\mathrm s}|\bar{\textbf{a}}_1\rangle 
+ \langle\bar{\textbf{a}}_1|\rho_{\mathrm s}|\bar{\textbf{a}}_1\rangle \big) 
\end{align}

By using the above relations, we obtain the following three equations. Since $|\textbf{a}_0\rangle = |\bar{\textbf{a}}_1\rangle$ and $|\textbf{a}_1\rangle = |\bar{\textbf{a}}_0\rangle$, all three equations yield identical values and reduce to the same expression for estimating the fidelity of the $n$-qubit GHZ state.
\begin{align}
P_{+}^{0}+P_{-}^{0}+P_{+}^{1}-P_{-}^{1}=\frac{1}{2}(\langle{\textbf{a}_0}|\rho_{\mathrm s}|\textbf{a}_0\rangle+\langle{\textbf{a}_1}|\rho_{\mathrm s}|\bar{\textbf{a}_1}\rangle+\langle{\bar{\textbf{a}}_1}|\rho_{\mathrm s}|\textbf{a}_1\rangle)+\langle{\bar{\textbf{a}}_0}|\rho_{\mathrm s}|\bar{\textbf{a}}_0\rangle)\\
2P_{+}^{0}=\frac{1}{2}(\langle\textbf{a}_0|\rho_{\mathrm s}|\textbf{a}_0\rangle+\langle\textbf{a}_0|\rho_{\mathrm s}|\bar{\textbf{a}}_0\rangle)+\langle\bar{\textbf{a}}_0|\rho_{\mathrm s}|\textbf{a}_0\rangle+\langle{\bar{\textbf{a}}_0}|\rho_{\mathrm s}|\bar{\textbf{a}}_0\rangle)\\
2P_{+}^{1}=\frac{1}{2}(\langle\textbf{a}_1|\rho_{\mathrm s}|\textbf{a}_1\rangle+\langle\textbf{a}_1|\rho_{\mathrm s}|\bar{\textbf{a}}_1\rangle)+\langle\bar{\textbf{a}}_1|\rho_{\mathrm s}|\textbf{a}_1\rangle+\langle{\bar{\textbf{a}}_1}|\rho_{\mathrm s}|\bar{\textbf{a}}_1\rangle)
\end{align}

In our experiment, we employed the combination $P_{+}^{0} + P_{-}^{0} + P_{+}^{1} - P_{-}^{1}$ to estimate the GHZ-state fidelity. This combination incorporates four measurement outcomes, thereby providing improved statistical robustness and reduced standard error compared to estimators based on fewer terms.

\begin{figure*}[t]
    \centering
    \includegraphics[width=0.88\textwidth]{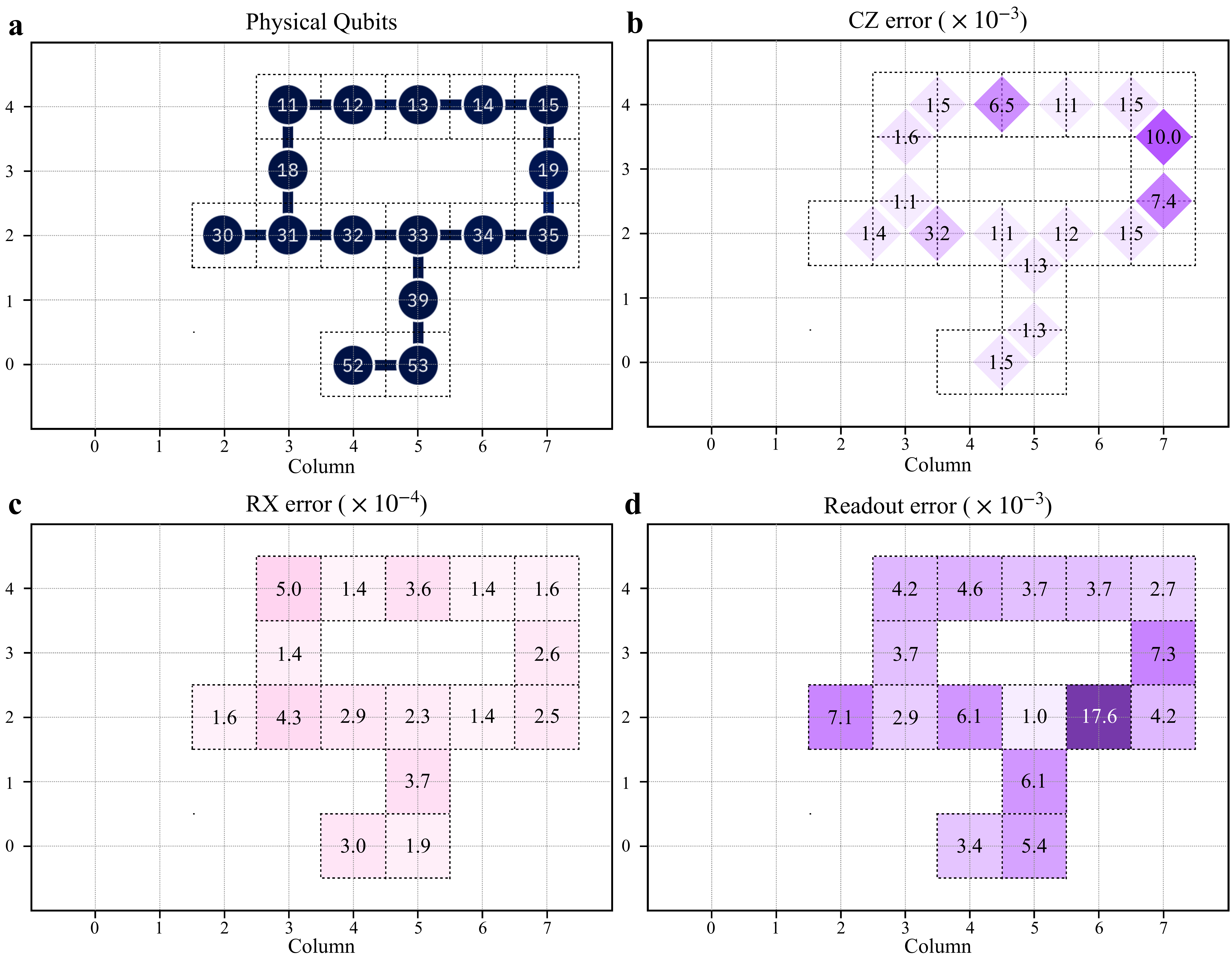}
    \caption{\small\justifying
Hardware configuration for GHZ-state fidelity estimation (up to $n$=15).
\textbf{a,} Physical qubits employed in the experiments with system sizes up to $n = 15$. All experiments were performed on the \textit{ibm\_aachen} device. For every configuration up to 15 qubits, qubit 31 was designated as the meter qubit. 
\textbf{b,} $CZ$ error rates between adjacent qubits. The average $CZ$ error rate was $2.94\times 10^{-3}$. Notably, the error rates for the couplings (12,13), (15,19), and (19,35) were significantly higher than those of the other couplings, which explains the slightly larger drops in fidelity observed when increasing the system size from $n=5$ to $n=6$, from $n=8$ to $n=9$, and from $n=9$ to $n=10$.
\textbf{c,} Single-qubit $RX$ gate error rates for each physical qubit. The average $RX$ error rate was $2.53\times 10^{-4}$, confirming that single-qubit gate errors are substantially lower than two-qubit gate errors.  
\textbf{d,} Readout error rates for the physical qubits. The average readout error was $5.23\times 10^{-3}$, which indicates that readout constituted the dominant error source. To mitigate this, we applied QREM throughout the experiments.  
}
    \label{fig:exp_dens}
\end{figure*}

\subsection{B. Experimental feature and layout}

\begin{figure*}[h!]
    \centering
    \includegraphics[width=0.88\textwidth]{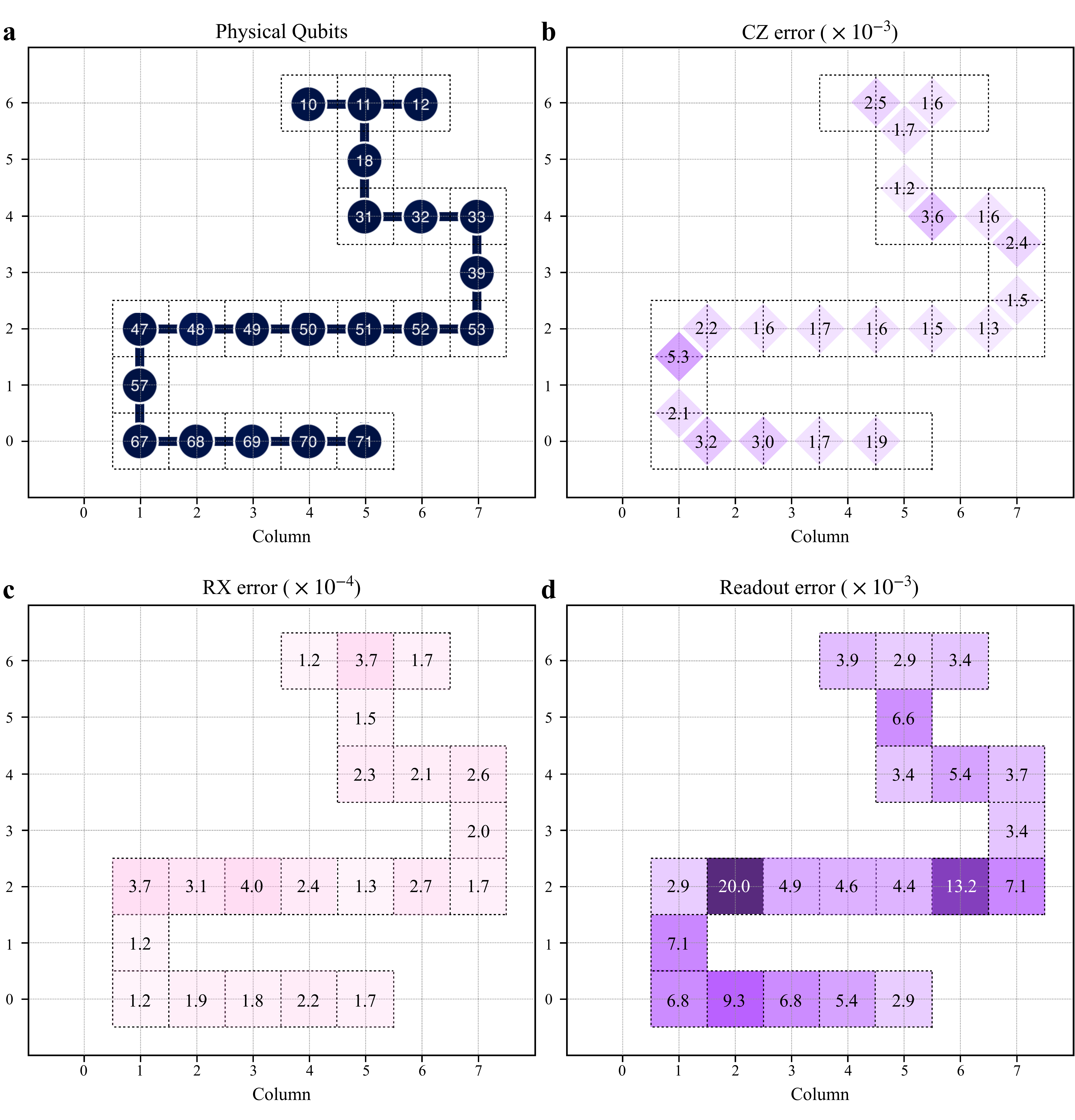}
    \caption{\small\justifying
Hardware configuration for GHZ-state fidelity estimation ($n$=20). 
Due to the change in calibration data after the $n=15$ experiment, we modified the qubit layout to utilize qubits with lower error rates.  
\textbf{a,} Physical qubits used for experiments with $n = 20$. For this experiment, we employed qubit 11 as the meter qubit.  
\textbf{b,} $CZ$ error rates between adjacent qubits. The average two-qubit $CZ$ error rate was $2.18\times10^{-3}$. In this configuration, no couplings exhibited exceptionally large error rates; the highest observed value was $5.3\times10^{-3}$ for the coupling (47,57), which is approximately half of the maximum value reported in Fig.~\ref{fig:exp_dens}\textbf{b} ($10.0\times10^{-3}$).  
\textbf{c,} Single-qubit $RX$ gate error rates for each physical qubit. The average $RX$ gate error rate was $2.18\times10^{-4}$, confirming that single-qubit gate errors are negligible compared to two-qubit gate errors.  
\textbf{d,} Readout error for the physical qubits. The average readout error was $6.10\times10^{-3}$, indicating that readout error is the dominant error source. As mentioned before, this effect was mitigated through QREM.  
}
    \label{fig:exp_dens_20}
\end{figure*}

The experimental configurations employed for GHZ-state fidelity estimation is presented. For system sizes up to $n=15$, we used the physical qubits shown in Fig.~\ref{fig:exp_dens}, along with their corresponding $CZ$, $RX$, and readout error rates. During the subsequent 20-qubit experiment, however, the device calibration had changed, resulting in increased error rates for the originally selected qubits. To mitigate the impact of these errors, we reconfigured the hardware by selecting an alternative set of qubits with lower error rates, as shown in Fig.~\ref{fig:exp_dens_20}. The corresponding physical qubit indices are summarized in Table~\ref{tab:physical qubits}.

The same qubit layout strategy was used as in Supplementary Note 2. For the state preparation of an $n$-qubit GHZ state, we entangled qubits $q_1$, $q_2$, $q_3$, $\dots$, $q_{n-3}$, $q_n$, and $q_{n-1}$, followed by $SWAP$ operations between $q_n$ and $q_{n-2}$. For matrix-element selection, we employed the same scheme illustrated in Fig.~\ref{fig:circuit layout}.

\begin{table*}[b!]
    \centering
    \includegraphics[width=0.9\textwidth]{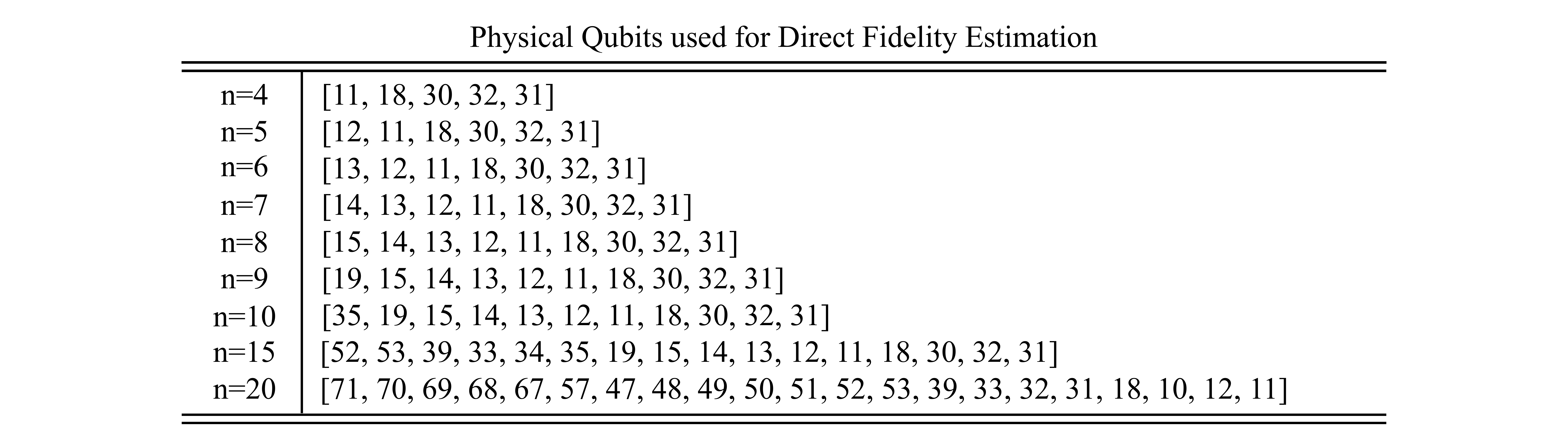}
    \caption{\small\justifying
Physical qubits used for GHZ-state fidelity estimation.  
Physical qubits corresponding to each logical qubit $q_1$ to $q_{n+1}$ are shown from left to right. 
}
    \label{tab:physical qubits}
\end{table*}

\subsection{C. Error mitigation and statistical analysis}

We provide details of the error mitigation methods and statistical analysis employed in GHZ-state fidelity estimation. As a representative case, we consider the $n=4$ experiment, for which the circuit layout is identical to that shown in Fig.~\ref{fig:circuit layout}, except that the physical qubits used were $\{11, 18, 30, 32, 31\}$. Across all experimental configurations, readout errors and two-qubit gate errors were identified as the dominant sources of noise, as shown in Figs.~\ref{fig:exp_dens} and~\ref{fig:exp_dens_20}. To suppress readout errors, we applied quantum readout error mitigation (QREM), as described in Supplementary Note~2. To mitigate the impact of two-qubit gate errors, we employed the zero-noise extrapolation (ZNE) technique~\cite{giurgica2020digital}, which enables estimation of ideal expectation values in the absence of noise originating from the two-qubit gates used for matrix-element selection.

ZNE was applied exclusively to the matrix-element selection operations to prevent modification of the target state, since in practical real-time applications the target state must remain intact. To ensure the effectiveness of ZNE, we employed Pauli twirling (PT) techniques~\cite{wallman2016randomized,hashim2020randomized}. PT was used to convert the physical noise channel into an effective Pauli noise channel, ensuring that the increased noise in the three-fold and five-fold circuits scales approximately linearly with respect to gate-folding depth. Here, ``1-fold,'' ``3-fold,'' and ``5-fold'' correspond to repeating the matrix-element selection once, three times, and five times, respectively.

\begin{figure*}[t]
    \centering
    \includegraphics[width=0.88\textwidth]{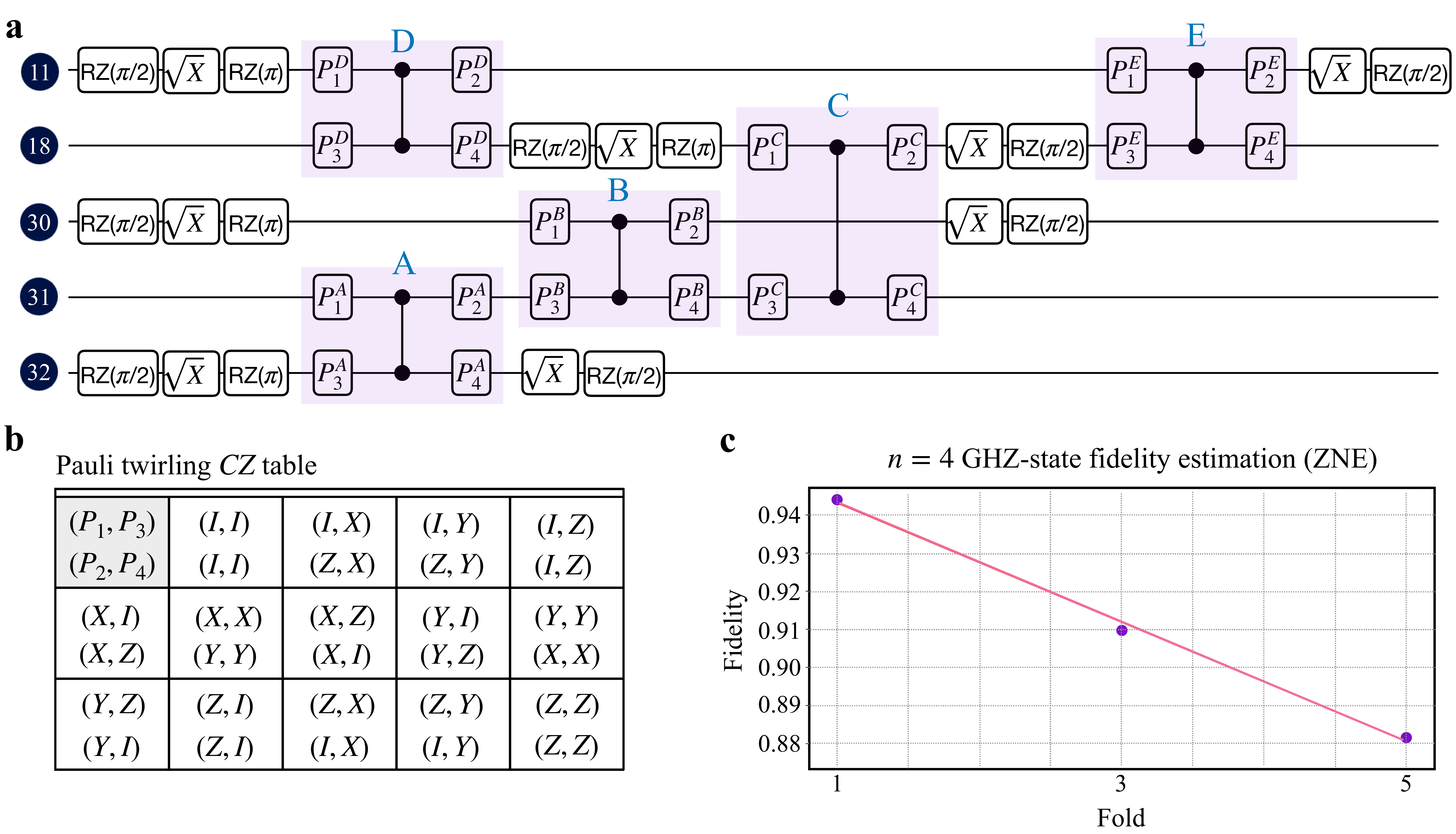}
    \caption{\small\justifying 
    Pauli twirling and zero-noise extrapolation scheme for GHZ-state fidelity estimation.
\textbf{a,} Pauli twirling applied to the transpiled matrix-element selection block in Fig.~\ref{fig:circuit layout}, using physical qubits $\{11, 18, 30, 31, 32\}$ with qubit 31 as the meter qubit. Since the native gates in the \textit{ibm\_aachen} are $CZ$, $RZ$, and $SX$, all $CNOT$ gates are transpiled into $CZ$ gates together with $RZ$ and $SX$ gates. To mitigate the effect of two-qubit errors in ZNE, random Pauli sets were inserted around every $CZ$ gate (shown in purple). Each set A, B, C, D, and E contains the four Pauli operators $P_1, P_2, P_3, P_4$. 
\textbf{b,} Pauli gate combinations that leave the $CZ$ gate invariant. These combinations were used to generate random Pauli sets by sampling from this table and randomly assigning them to the sets A–E.  
\textbf{c,} Example of $n=4$ GHZ-state fidelity estimation with ZNE. The plot shows the linear scaling between the 1-fold, 3-fold, and 5-fold circuits when Pauli twirling is applied, enabling ZNE to extrapolate the zero-noise value, which in this case is $0.959$.
}
    \label{fig:error mitigation}
\end{figure*}

Figure~\ref{fig:error mitigation}\textbf{a} shows the transpiled circuit for the $n=4$ matrix-element selection implementing ${U}_{\mathrm{ES}}^{\mathbf{1}} = {X}^{\otimes 4}$. Pauli twirling (PT) was applied to all native two-qubit ($CZ$) gates within this block, as illustrated in Fig.~\ref{fig:error mitigation}\textbf{a}. The Pauli operator sets used for twirling each $CZ$ gate are shown in Fig.~\ref{fig:error mitigation}\textbf{b}.

For the $n=4$ case, the projector selection block consists of five sets of $CZ$ gates, labeled A, B, C, D, and E. For each $CZ$ gate, a set of Pauli operators $(P_1, P_2, P_3, P_4)$ was selected from the CZ Pauli table, satisfying the following condition:
\begin{equation}
(P_1 \otimes P_3) \text{CZ}(P_2 \otimes P_4) = \text{CZ}.
\label{eq:CZ}
\end{equation}
For each experiment, we randomly selected Pauli sets for (A, B, C, D, E) and executed the circuit with 1000 shots. This procedure was repeated 100 times with different random Pauli sets, resulting in a total of 100,000 shots, since Pauli twirling requires averaging over random Pauli realizations.

For statistical analysis, we used bootstrapping. From the 100 random Pauli sets, we generated new `bootstrap sets' of 100 elements by sampling with replacement. This procedure was repeated 50 times, producing 50 bootstrap sets, each containing 100 random Pauli realizations. We then computed the GHZ-state fidelity for each Bootstrap set and calculated the mean fidelity and standard error across the 50 Bootstrap sets. This process yielded the Pauli-twirled mean fidelity and standard error for the 1-fold circuit. Then we constructed 1-fold, 3-fold, and 5-fold circuits and applied the same PT and Bootstrap procedures described above. The mean fidelity values for these three noise levels were then plotted in Fig.~\ref{fig:error mitigation}, and a linear fit was applied to extrapolate the zero-noise value, resulting in an estimated fidelity of $0.959$. This procedure was consistently used in all GHZ-state fidelity estimation experiments to obtain the ``with ZNE'' results reported in the main text.

\end{document}